\documentclass[journal,comsoc]{IEEEtran}

\usepackage{hyperref}
\usepackage{cite}
\usepackage{url}
\usepackage{graphicx}
\usepackage{epstopdf}
\usepackage{amssymb}
\usepackage{amsmath}
\usepackage{multirow}
\usepackage{longtable}
\usepackage{multicol}
\usepackage{subfigure}
\usepackage[ruled,vlined]{algorithm2e}
\usepackage{bbm}
\usepackage{array}
\usepackage{empheq}
\usepackage{dblfloatfix}
\usepackage{placeins}
\usepackage{color}
\usepackage[normalem]{ulem}
\usepackage{amsthm}
\usepackage{booktabs}
\usepackage[flushleft]{threeparttable} 

\newcolumntype{?}{!{\vrule width 1pt}}

\begin{document}
\title{\huge Ultrareliable and Low-Latency Communication Techniques for Tactile Internet Services \\}
\author{\normalsize Kwang~Soon~Kim$^{*\dag}$,~Dong~Ku~Kim$^{*}$,~Chan-Byoung~Chae$^{**}$,~Sunghyun~Choi$^{+}$, Young-Chai~Ko$^{++}$,\\~Jonghyun~Kim$^{*}$,~{Yeon-Geun}~Lim$^{**}$,~Minho~Yang$^{*}$,~Sundo~Kim$^{+}$, Byungju~Lim$^{++}$,~Kwanghoon~Lee$^{*}$, and~Kyung~Lin~Ryu$^{*}$  
\thanks{\small
This work was supported by the Institute for Information \& communications Technology Promotion (IITP) grant funded by the Korea government (MSIT) (2015-0-00300, Multiple Access Technique with Ultra-Low Latency and High Efficiency for Tactile Internet Services in IoT Environments)}
\thanks{\small $^{*}$ The authors are with the School of Electrical and Electronic Engineering, Yonsei University, Korea.}
\thanks{\small $^{**}$ The authors are with the School of Integrated Technology, Yonsei University, Korea.}
\thanks{\small $^{+}$ The authors are with the Department of Electrical and Computer Engineering, Seoul National University, Korea.}
\thanks{\small $^{++}$ The authors are with the School of Electrical Engineering, Korea University, Korea.}
\thanks{\small $^\dag$: Corresponding author (ks.kim@yonsei.ac.kr)}
}
\markboth{Kim {\it et al.}: Multiple Access Techniques with Ultra-Reliability and Low-Latency for Tactile Internet Services}{manuscript submitted to Proceedings of IEEE} 
\maketitle
\vspace{-1cm}
\begin{abstract}
This paper presents novel Ultrareliable and low-latency communication (URLLC) techniques for URLLC services, such as Tactile Internet services. Among typical use-cases of URLLC services are tele-operation, immersive virtual reality, cooperative automated driving, and so on. In such URLLC services, new kinds of traffic such as haptic information including kinesthetic information and tactile information need to be delivered in addition to high-quality video and audio traffic in traditional multimedia services. Further, such a variety of traffic has various characteristics in terms of packet sizes and data rates with a variety of requirements of latency and reliability. Furthermore, some traffic may occur in a sporadic manner but require reliable delivery of packets of medium to large sizes within a low latency, which is not supported by current state-of-the-art wireless communication systems and is very challenging for future wireless communication systems. Thus, to meet such a variety of tight traffic requirements in a wireless communication system, novel technologies from the physical layer to the network layer need to be devised. In this paper, some novel physical layer technologies such as waveform multiplexing, multiple access scheme, channel code design, synchronization, and full-duplex transmission for spectrally-efficient URLLC are introduced. In addition, a novel performance evaluation approach, which combines a ray-tracing tool and system-level simulation, is suggested for evaluating the performance of the proposed schemes. Simulation results show the feasibility of the proposed schemes providing realistic URLLC services in realistic geographical environments, which encourages further efforts to substantiate the proposed work \footnote{Readers are invited to visit \url{https://www.dropbox.com/s/8qnjzla9g2lbr4e/Tactile-2017-SLS-Implementation.mp4?dl=0} for a video clip introducing the proposed work.}. 
\end{abstract}
\newpage
\section{Introduction}

\IEEEPARstart{O}{wing} to continuously increasing demands for new vertical services, academia and industry have been placing a huge emphasis on developing the fifth generation (5G) enabling technologies. 
According to the international mobile telecommunication (IMT) vision for 2020 \cite{IMT2020}, emerging 5G services include enhanced mobile
broadband services, massive Internet of Things (IoT), and Ultrareliable and low-latency
communication (URLLC) services, such as Tactile Internet services\footnote{In this paper, URLLC service and Tactile Internet service are used interchangeably because both share most of interesting use-cases in literature and pursue the same service vision and requirements.}. Among them, URLLC services are considered as the most challenging applications in 5G or future cellular systems, and their typical use cases include collaborative automated cars, tele-operations, interpersonal communications (ICs), and immersive virtual reality (IVR) services \cite{ehealth,automation,industry,ind2,nokia,VKKV,PHCR}. 

Unlike the classical high-quality multimedia streaming service, in which high-rate information flows from a source to a sink, sensing information, control and command information, and feedback information that occurred by the actuation according to the control and command form a loop for an information flow in typical URLLC services. In a typical tele-surgery example \cite{telesurgery1,telesurgery2}, a real-time high-quality video sensing information of the affected area of a patient needs to be delivered to a surgeon and the surgeon controls a remote surgical robot, wherein elaborate kinesthetic information of the surgeon's hands and fingers needs to be delivered to the robot; the force and tactile sensing information that occurred from the interaction between the robot and the affected area needs to be fed back to help the surgeon along with the video sensing information. Note that such a typical information flow requires a delivery of high-rate information up to several hundred megabits per second (Mbps) with and end-to-end latency as low as $1$ ms and a reliability as high as $99.9999999\%$ (i.e., packet error rate of $10^{-9}$), which reveals some challenging aspects of developing URLLC enabling technologies in cellular communication systems.

Studies on typical URLLC use-cases have been carried out, in which typical traffic characteristics and quality of service (QoS) requirements of some URLLC use-cases are reported. In \cite{ehealth}, a tele-surgery use case is described in which audio, video, and haptic information needs to be delivered within an end-to-end latency as low as $1$ ms and with an extremely high reliability (block error rate (BLER) down to $10^{-9}$). In \cite{automation}, intelligent transportation examples are described, such as cooperative collision avoidance and high-density platooning, in which sensing information needs to be exchanged within an end-to-end latency as low as $5$ ms and with high reliability (frame error rate (FER) down to $10^{-6}$). Further, in \cite{industry,ind2,nokia}, industry automation examples are described, such as time-critical process optimization inside factory and remote control, in which video, audio, and haptic information needs to be delivered within a sub millisecond end-to-end latency and with an extremely high reliability (BLER down to $10^{-9}$). Recently, the IEEE standardization activity on Tactile Internet (IEEE P1918.1) was launched, in which Tactile Internet architecture, functional entities, and various use-cases have been investigated. In \cite{VKKV,PHCR}, detailed traffic characteristics of video, audio, and haptic information such as packet size, arrival rate, and arrival model with QoS such as latency and reliability requirements are described. 
These examples and scenarios show that traffic characteristics of typical URLLC services can be quite various in terms of their packet sizes and arrival models and their QoS requirements can be quite extreme \cite{Fettweis,Steinbach}; therefore, these aspects should be taken into account when developing URLLC techniques.

To support such low-latency requirements of URLLC services, studies in the 3rd Generation Partnership Project (3GPP) on the current long-term evolution (LTE) systems have been performed \cite{3gpplow}, in which typical downlink (DL)/uplink (UL) radio access and handover latencies are reported as $17$/$7.5$ ms and $50$ ms, respectively, and the transmit-time-interval (TTI) reduction, processing time reduction, semi-persistent scheduling, and grant-free access are enumerated as possible remedies.
In addition, many technical aspects in the cellular network, including waveform numerology such as symbol length and subcarrier spacing, frame structure, multiple access scheme, pilot design, link adaptation strategy, and scheduling policy need to be designed carefully for URLLC \cite{Ericsson,IntelCorp,Qualcomm}.
3GPP is standardizing a new radio interface for 5G as the new radio (NR), aiming to reduce DL/UL radio access latency to 0.5 ms \cite{NRScenario}.
In \cite{NRPHY,NRProtocol}, scalable subcarrier spacing parameters for shorter orthogonal frequency division multiplexing (OFDM) symbol length and mini-slots comprised of various number of OFDM symbols (1-13) are adopted for implementing short TTIs. Further, various ideas on URLLC and enhanced mobile broadband (eMBB) multiplexing for efficient resource utilization \cite{DLmux1,DLmux2,DLmux3} and various ideas on two-way grant-based and grant-free multiple access proposals \cite{gfma1,gfma2,gfma3} to reduce the uplink protocol latency have been discussed. 
However, such simple suggestions on providing low-latency protocols and frame structures should be just the beginning, as practical URLLC services need a simultaneous provision of low-latency and ultra-reliability with high spectral efficiency, which is very challenging. 

In a cellular system, the channel impulse (or frequency) responses of wireless fading channels are not fully predictable and the fluctuation on the received signal-to-noise-plus-interference ratio (SINR) is one of the most challenging aspects for reliable information delivery. The current 3GPP long-term evolution (LTE) employs an appropriate scheduling to utilize multiuser diversity, adaptive modulation and coding (AMC) according to channel quality information measured at a receiver, and hybrid automatic repeat and request (HARQ) for an efficient retransmission to provide high reliability as well as high spectral efficiency \cite{LTEbook}. However, such an approach requires delays for channel quality measure and feedback, scheduling, and retransmission that it becomes inappropriate for delivering highly latency-sensitive information, although it is the most efficient way to deliver latency-insensitive information. Although some diversity schemes and fast HARQ schemes for better reliability at a low-latency are considered, such as in \cite{diver,harq,harq2}, their reliability levels and the resulting spectral efficiencies are far from what is required for practical URLLC services. Further, the design of the physical (PHY) layer and medium access control (MAC) layer technologies for URLLC need to consider the variety of different traffic characteristics and the QoS of URLLC services.

Since 2015, the authors had formed a joint URLLC research team and focused on developing spectrally-efficient protocol and multiple access technologies that guarantee both tight low-latency and ultra-reliability requirements for URLLC. To provide ultra-reliability, a large amount of diversity obtained from large degrees of freedom is essential, especially without either instantaneous channel quality feedback or retransmissions in fading channels. Thus, considering a large-scale antenna system (LSAS) (or massive multiple-input multiple-output (MIMO) system) is a natural consequence \cite{MarzettaNoncooperative,Lim2015}. In this paper, some novel multiple access schemes for URLLC based on the LSAS are introduced and waveform multiplexing and full-duplex communication techniques are also introduced to further enhance the spectral efficiency and reduce the latency. In addition, a new evaluation methodology is introduced by combining a system-level simulator and a ray-tracing tool with digital maps on real environments and the performance evaluation results are provided.

\section{Some Use Cases and Traffic Requirements for Tactile Internet Services} \label{S2}

\subsection{Some Use Cases} \label{2A}

\footnotesize
\begin{table*}[]
	\centering
	\caption{ Typical Traffic Characteristics and QoS for Some Use-cases }
	\label{table1}
	\begin{threeparttable}
		\begin{tabular}{|c|c|c|c|c|c|}
			\hline
			\textbf{Application} & \textbf{Types} & \textbf{\begin{tabular}[c]{@{}c@{}} Reliability ($R$) \end{tabular}} & \textbf{\begin{tabular}[c]{@{}c@{}} Typical Air-latency ($L$) \end{tabular}} & \textbf{Burst size ($B$)} & \textbf{Arrival model ($A$)} \\ \hline
			\multirow{3}{*}{\begin{tabular}[c]{@{}c@{}}IVR\end{tabular}} & Haptics & \begin{tabular}[c]{@{}c@{}}99.9\%\tnote{1},\\$>99.999$\%\tnote{2}\end{tabular} & 0.5-2~ms & \begin{tabular}[c]{@{}c@{}} 2-8~B/DoF\\ (1/10/100/1000~DoFs)\end{tabular} & \begin{tabular}[c]{@{}c@{}}1000-5000~pkt/s (P)\tnote{1},\\ 100-500~pkt/s (GE)\tnote{2}\end{tabular} \\ \cline{2-6} 
			& Video & $>99.999$\% & 0.5-2~ms & 1-30~KB & 100-1000~pkt/s (P) \\ \cline{2-6} 
			& 3D Audio & 99.9\% & 0.5-2~ms & 100~B & 10-1000~pkt/s (P) \\ \hline
			\multirow{3}{*}{\begin{tabular}[c]{@{}c@{}}Tele-operation\\ (T)\end{tabular}} & Haptics & \begin{tabular}[c]{@{}c@{}}99.9\%\tnote{1},\\ $>99.999$\%\tnote{2}\end{tabular} & \begin{tabular}[c]{@{}c@{}}0.5-2~ms (high-dynamic)\\ 10~ms (dynamic) \\ 100~ms (static)\end{tabular} & \begin{tabular}[c]{@{}c@{}} 2-8~B/DoF\\ (1/10/100/1000~DoFs)\end{tabular} & \begin{tabular}[c]{@{}c@{}}1000-5000~pkt/s (P)\tnote{1},\\ 100-500~pkt/s (GE)\tnote{2}\end{tabular} \\ \cline{2-6} 
			& Video & $>99.999$\% & 5~ms & 1-10~KB & 100-1000~pkt/s (P) \\ \cline{2-6} 
			& Audio & 99.9\% & 5~ms & 50-100~B & 10-1000~pkt/s (P) \\ \hline
			\multirow{4}{*}{\begin{tabular}[c]{@{}c@{}}Automotive\\ (A)\end{tabular}} & Haptics & \begin{tabular}[c]{@{}c@{}}99.9\%\tnote{1},\\     $>99.999$\%\tnote{2}\end{tabular} & \multirow{4}{*}{\begin{tabular}[c]{@{}c@{}}0.5-2~ms (life-critical)\\     10~ms (dynamic)\\ 100~ms (static)\end{tabular}} & \begin{tabular}[c]{@{}c@{}}2-8~B/DoF\\ (1/10/100~DoFs)\end{tabular} & \begin{tabular}[c]{@{}c@{}}1000-5000~pkt/s (P)\tnote{1},\\ 100-500~pkt/s (GE)\tnote{2}\end{tabular} \\ \cline{2-3} \cline{5-6} 
			& Sensor & $>99.999$\% &  & 1-5~KB & 100-1000~pkt/s (E)\tnote{3} \\ \cline{2-3} \cline{5-6} 
			& Video & 99.9\% &  & 1-10~KB & 100-1000~pkt/s (P) \\ \cline{2-3} \cline{5-6} 
			& Audio & 99.9\% &  & 50-100~B & 10-1000~pkt/s (P) \\ \hline
			\multirow{5}{*}{\begin{tabular}[c]{@{}c@{}} IoD \end{tabular}} & Haptics & \begin{tabular}[c]{@{}c@{}}99.9\%\tnote{1},\\     $>99.999$\%\tnote{2}\end{tabular} & \begin{tabular}[c]{@{}c@{}} 0.5-2~ms (kinesthetic)\\10~ms (tactile)\end{tabular} & \begin{tabular}[c]{@{}c@{}}2-8~B/DoF\\ (1/10/100~DoFs)\end{tabular} & \begin{tabular}[c]{@{}c@{}}1000-5000~pkt/s (P)\tnote{1},\\ 100-500~pkt/s (GE)\tnote{2}\end{tabular} \\ \cline{2-6} 
			& GPS & 99.9\% & 10~ms & 2~KB & 100-1250~pkt/s (P) \\ \cline{2-6} 
			& Sensor & $>99.999$\% & 10~ms & 1-5~KB & 100-1000~pkt/s (E)\tnote{3} \\ \cline{2-6} 
			& Video & $>99.999$\% & 1-10~ms & 1-20~KB & 100-1000~pkt/s (P) \\ \cline{2-6} 
			& Audio & 99.9\% & 1-10~ms & 50-100~B & 10-1000~pkt/s (P) \\ \hline
			\multirow{3}{*}{\begin{tabular}[c]{@{}c@{}} HIC \end{tabular}} & Haptics & \begin{tabular}[c]{@{}c@{}}99.9\%\tnote{1},\\     $>99.999$\%\tnote{2}\end{tabular} & \begin{tabular}[c]{@{}c@{}}1-2~ms (interaction)\\ 10-100~ms (observation)\end{tabular} & \begin{tabular}[c]{@{}c@{}}2-8~B/DoF\\ (1/10/100/1000~DoFs)\end{tabular} & \begin{tabular}[c]{@{}c@{}}1000-5000~pkt/s (P)\tnote{1},\\ 100-500~pkt/s (GE)\tnote{2}\end{tabular} \\ \cline{2-6} 
			& Video & $>99.999$\% & 5~ms & 1-30~KB & 100-1000~pkt/s (P) \\ \cline{2-6} 
			& Audio & 99.9\% & 5~ms & 50-100~B & 10-1000~pkt/s (P) \\ \hline
		\end{tabular}
		\begin{tablenotes}
			\item $^1$w/o compression, P: Periodic ~ $^2$w/ compression, GE: Gilbert-Elliot (such as in \cite{GE}) ~$^3$ E: Event-driven (Sporadic)
		\end{tablenotes}
	\end{threeparttable}
\end{table*}
\normalsize

\subsubsection{Immersive Virtual Reality}

The immersive virtual reality (IVR) technology allows people to use their senses to interact with virtual entities in remote or virtually created environments such that they can perceive all five senses when they are in such remote or virtual environments \cite{PHCR}. Because of  its interaction capability beyond the physical limitation, it has been drawing great interest in industries such as gaming, education, and health care \cite{IntelIVR}.

Among the five senses, the vision, sound, and touch senses represent the primary focus and their traffic types and characteristics can be categorized according to each sense. For vision sensing, since the motion-to-photon latency should be within $10$-$20$ ms, the allowed air latency ranges from sub milliseconds to a few milliseconds \cite{nokia,QualIVR}. Further, considering the field of view, be it in three dimensions, or extremely high definition (32K) \cite{HuaweiIVR}, a required data rate for vision information would be in the range between $10$ Mbps to $1$ Gbps with $99.9\%$-$99.999\%$ reliability. For audio sensing, the audio information includes not only high-fidelity sound but also considerations for three-dimensional head rotations. For touch sensing, haptic information exchange is required, in which tactile information comprising several bytes for each degree of freedom (DoF) \cite{PHCR} times the number of DoFs (i.e., the number of touch spots) up to thousands and the kinesthetic information comprising several bytes per each DoF \cite{PHCR} times the number of DoFs (i.e., the number of joints in the human body) up to hundreds with $99.999\%$ reliability. 

\subsubsection{Tele-operation}

Tele-operations, such as tele-surgery,  tele-maintenance, and tele-soccer using remote robotic avatars, allow people to control slave devices such as robots in distant or inaccessible environments to perform complex tasks \cite{PHCR,Tele.ProcIEEE}. The exchange of haptic information, such as force, torque, velocity, vibration, touch, and pressure, is required between the master and slave devices, and the delivery of high-quality video and audio information is required from the slave devices to the master devices \cite{PHCR}.

The required data rates and latency requirements of the traffic for tele-operation vary according to the required control precisions for slave devices and the dynamics of remote environments where the slave devices are placed. In a highly dynamic environment such as the one reported in \cite{robocup}, haptic information exchange should be within a few milliseconds such that the allowed air latency is less than or equal to one millisecond. Further, for applications requiring extremely high control precision such as in \cite{telesurgery1,telesurgery2}, the delivery of very high-rate video information and the exchange of delicate haptic information with reliability higher than $99.999\%$ is required. Further, for remote skill training such as in \cite{MDohlerNews}, the number of DoFs can be hundreds to thousands.

\subsubsection{Automotive and Internet of Drones (IoD)}

Future cars need connectivity with the infrastructure and other cars for collaborative autonomous driving and in-car entertainments \cite{automation}. Therefore, a large amount of sensing information needs to be exchanged in a very low latency. Similar to tele-operation applications, the required latency depends on the dynamics of neighboring environments such that the allowed air latency can be less than or equal to one millisecond with high reliability. In addition, for collaborative autonomous driving using artificial intelligence, high-quality video and audio information exchange with high reliability among neighboring cars may be required. In remote driving, haptic information exchange with DoFs up to several tens to hundreds may be required \cite{PHCR}.

Applications using unmanned aerial vehicles (UAVs), such as drones, are also emerging and among them are drones for public safety, remote explorations, logistics, flying base stations, etc. \cite{PHCR,UAV1}. Owing to the high dynamics in such UAV environments, real-time video, audio, and haptic information should be exchanged within a low latency, i.e., the allowed air latency less than or equal to one millisecond for kinesthetic information and a few milliseconds to tens of milliseconds for high-quality video/audio information and haptic information. 

\subsubsection{Interpersonal Communication}

Interpersonal communication (IC) supports the co-presence of distant users for social development or emotional interaction, and haptic IC (HIC) can deliver human touch as well, allowing for promising applications such as social networking, gaming, education, and training \cite{PHCR,IC1,IC2}.

High-quality video and audio information exchange is required with high reliability, similar to the IVR case, and haptic information exchange is also required. In static dialoguing, a low latency is required for the highly dynamic interaction of haptic information such that the allowed air latency can be as low as a few milliseconds \cite{PHCR}.

\subsection{Traffic Classification}

\footnotesize
\begin{table*}[]
	\centering
	\caption{Traffic Classification Example}
	\label{table2}
	\begin{tabular}{|c|c|c|c|c|c|}
		\hline
		\textbf{Class} & \textbf{Reliability ($R$)} & \textbf{\begin{tabular}[c]{@{}c@{}} Typical air-latency ($L$) \end{tabular}} & \textbf{Burst size ($B$)} & \textbf{Arrival model ($A$)} & \textbf{Applications} \\ \hline
		1 & 99.9-99.99999\% & $>50$~ms & 1-10~KB & \begin{tabular}[c]{@{}c@{}}10-5000~pkt/s (P),\\ 100-500~pkt/s (GE),\\ 100-1000~pkt/s (E)\end{tabular} & T, A, HIC \\ \hline
		2 & 99.9-99.99999\% & 10-50~ms & 1-20~KB & \begin{tabular}[c]{@{}c@{}}10-5000~pkt/s (P),\\ 100-500~pkt/s (GE),\\ 100-1000~pkt/s (E)\end{tabular} & T, A, HIC \\ \hline
		3 & 99.9-99.999\% & 2-10~ms & 1-30~KB & 10-5000~pkt/s (P) & IVR, T, A, IoD, HIC \\ \hline
		4 & 99.99999\% & 2~ms & 80~B & 100-50000~pkt/s (GE) & T, A, HIC \\ \hline
		5 & 99.999\% & 1~ms & 800~B & 10-5000~pkt/s (GE) & T, A, HIC \\ \hline
		6 & 99.99999\% & 2~ms & 5~KB & 100-1000~pkt/s (E) & A, IoD \\ \hline
		7 & 99.999\% & 2~ms & 8~KB & 100-500~pkt/s (E) & IVR, IoD \\ \hline
		8 & 99.999\% & 0.5~ms & 5~KB & 100-500~pkt/s (E) & IVR, T, A, IoD \\ \hline
	\end{tabular}
\end{table*}
\normalsize

In Table \ref{table1}, typical traffic characteristics of the use-cases in Section \ref{2A} are summarized, in which the traffic characteristics and QoS are represented by the typical air latency, target reliability, packet size, packet arrival rate and model. Here, the baseline of the traffic characteristics and QoS come from \cite{PHCR} but it is further assumed that typical air-latency requirements are set to 20\% of the corresponding end-to-end latency requirements and more DoFs and larger packet sizes up to ten times are expected in near future.

The current state-of-the-art cellular communication technology can support various traffic with different characteristics and QoS with good reliability and high spectral efficiency if the required latency is not so tight by controlling the radio resource control (RRC) connectivity of each user according to its activity, scheduling active users with good channel quality, AMC according to its channel quality, and retransmissions using HARQ.
However, extremely low latency and high reliability requirements of URLLC services necessitate classifying such traffic and operating different protocols and multiple access strategies according to different target latency and reliability levels.
Further, traffic characteristics such as the arrival model and rate also need to be taken into account to design such protocols and multiple access strategies.

First, traffic with loose latency requirements (i.e., $L>50$ ms) can be easily supported using a legacy strategy: radio-resource efficient LTE-style four-way RRC connection, scheduling, AMC, and HARQ regardless of traffic type, data rate, arrival model, and target reliability level. In this case, it is not difficult to satisfy the latency and reliability requirements and a higher spectral efficiency is of the most interest. 
When a latency requirement is slightly tight (i.e., $L$ is approximately tens of milliseconds), it is better to deal with such packets differently from those with loose latency requirements.
Some good approaches include reducing the number of handshakes in the RRC connection protocol as in \cite{gfma1,gfma2} and shortening the TTI as in \cite{NRPHY,NRProtocol}.
As the target latency is not so tight, it is possible to apply a grant-based multiple access with a radio resource management and a few retransmission can be allowed to guarantee the reliability requirement. 

As the latency requirement becomes tighter (i.e., $L$ is approximately several milliseconds), more elaborately designed techniques need to be applied and the traffic arrival model becomes important. For periodically generated packets, a semi-persistent scheduling can be applied to reserve the radio resources for such packets. Further, at least one or two retransmissions may be allowed such that a reliability requirement can be met with an LTE-style spectrally efficient radio resource management. However, in cases of bursty or sporadically generated packets, a grant-free multiple access, similar as in \cite{KimGFMA}, is necessary and it is important to guarantee a target reliability, which is very challenging. As users can transmit without any grant, the number of users sharing the same radio resources (i.e., a subchannel) varies and it becomes worse if traffic with different characteristics and QoS (such as packet size and target reliability level) are allocated in the same subchannel of a multiple access scheme. In addition, as the latency requirement becomes extremely tight (i.e., $L$ is less than or equal to $1$ ms), retransmission may not be allowed and it becomes very difficult to satisfy a reliability requirement at a reasonable spectral efficiency. A good approach is to classify traffic classes according to the characteristics and QoS such that packets with similar characteristics and QoS are allocated in each subchannel of a multiple access scheme.

In Table \ref{table2}, traffic in Table {\ref{table1} is classified as an example, mainly according to the latency and reliability requirements, as discussed in paragraphs above. Here, the first row represents a class with loose latency requirements, the second row represents a class with medium latency requirements, and the third row represents a class with low latency requirements but the packets are generated periodically. The other five classes represent very low latency requirements with bursty or sporadic packet arrival characteristics.
As they need to be served using a grant-free multiple access, those packets should be further classified according to latency, reliability requirements, and packet sizes so that traffic with similar characteristics are allocated to a subchannel for a reasonable spectrally-efficient radio resource management.

\section{Multiple Access Strategy for URLLC}

\begin{figure*} 
	\centering
	\includegraphics[width=.6\textwidth]{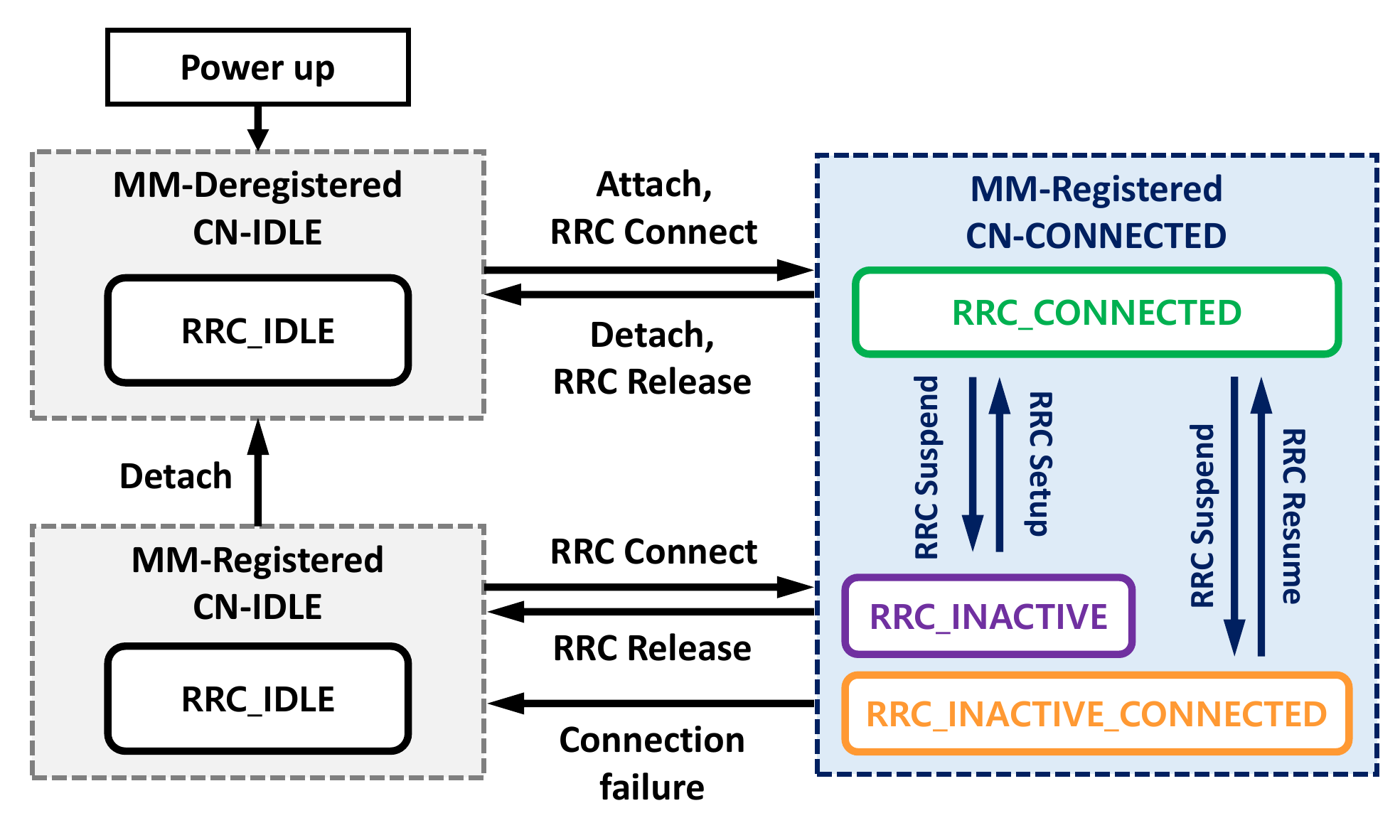} 
	\caption{RRC state transition diagram.} 
	\label{fig-uerrcstate}
\end{figure*}

In Section \ref{S2}, typical URLLC use-cases and traffic characteristics are introduced, which necessitate the development of not only a new frame structure with short TTIs and protocol concepts such as in \cite{NRScenario,NRPHY,NRProtocol}, but also elaborately designed strategies for user RRC state control, radio resource management and optimization, and novel multiple access techniques each suitable for the various traffic characteristics and QoS of URLLC services. 

In this section, a new user RRC control strategy is suggested with new states for serving traffic with low-latency requirements and the corresponding RRC connection protocols are suggested, in which different levels of protocol procedures, core network connection strategies, and radio resource allocation strategies are provided according to the traffic classes of URLLC users. In addition, DL and UL radio resources are appropriately partitioned to support different multiple access schemes, where each multiple access component handles traffic with similar characteristics and QoS for better spectral efficiency. To provide a high-level of reliability even in cases of extremely low latency requirements, an LSAS is assumed for a base station and a latency-optimal radio resource management scheme is suggested. According to each user's RRC state, a different level of radio resource allocation is provided to enhance the spectral efficiency while guaranteeing the latency and reliability requirements. 

\subsection{RRC Connection Protocols}

\begin{figure*}[t] 
	\centering
	\includegraphics[width=\textwidth]{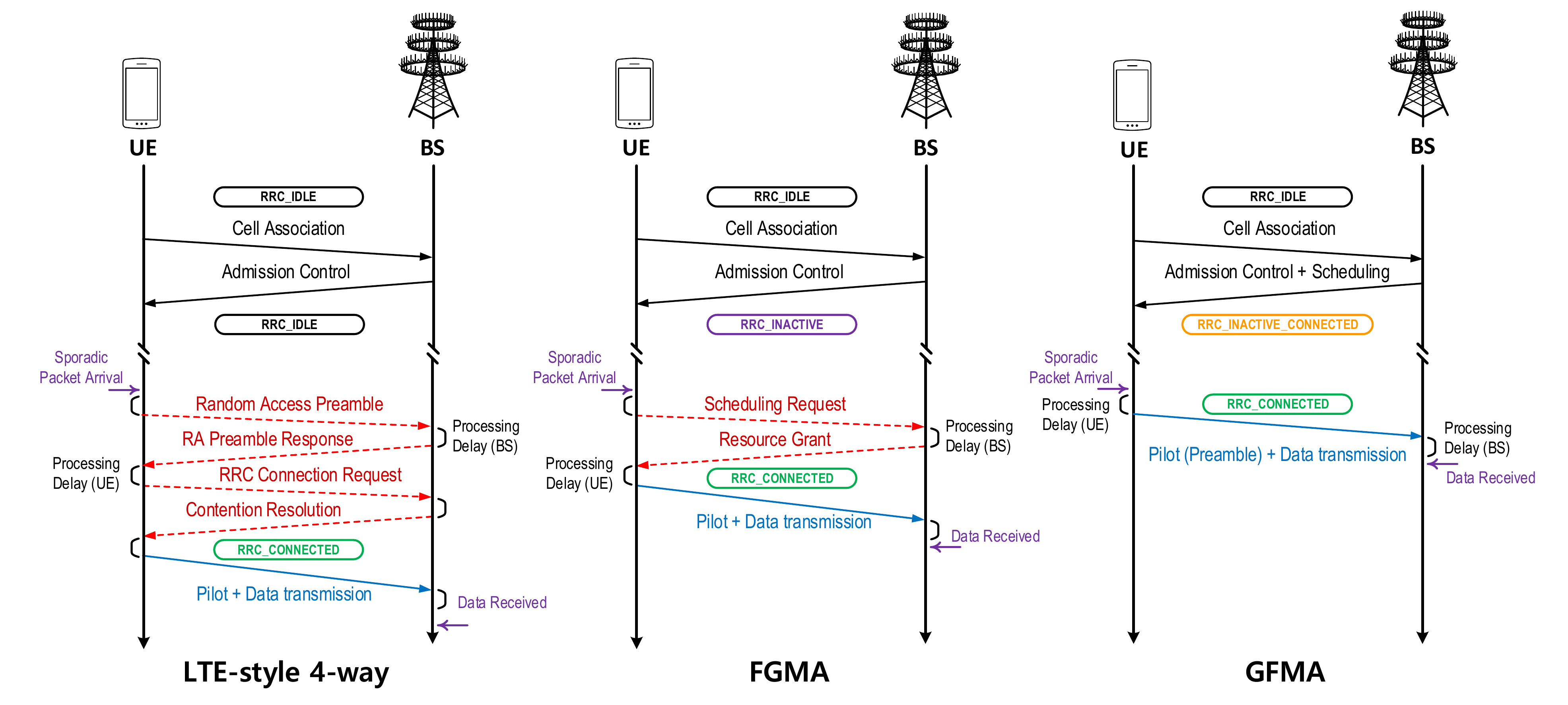}
	\caption{Three different procedures for RRC connection.} 
	\label{fig-procedure} 
\end{figure*}

Recently, in the 3GPP standardization for NR, a new user RRC state, {RRC\_INACTIVE}, has been defined \cite{RRCinactive_Nokia1,RRCinactive_Nokia2,RRCinactive_Nokia3}. According to a general description in \cite{RRCinactive_5GPPP}, {RRC\_INACTIVE} is different from {RRC\_IDLE} in that a user keeps the previous configuration information when suspended from {RRC\_CONNECTED}, so that it can resume RRC connectivity without a long delay. Further, from a core network's perspective, the {RRC\_INACTIVE} and  {RRC\_CONNECTED} are the same because core network is connected (i.e., {CN\_CONNECTED}) in both cases. If the RRC connectivity is lost, a user needs to perform the RRC connection setup similarly from {RRC\_IDLE}. 

In this paper, such a new state is further classified into two different states according to the required number of handshakes between base stations and users and their levels of allocated radio resources. As shown in Fig. \ref{fig-uerrcstate}, two new states, RRC\_INACTIVE and RRC\_INACTIVE \_CONNECTED are introduced.
Here, from a core network's perspective, both states are the same to the RRC\_CONNECTED.
To a user in RRC\_INACTIVE or RRC\_INACTIVE\_CONNECTED, preambles are allocated as dedicated radio resources in addition to the RRC configuration information and they uniquely indicate each user's identity and intended traffic classes.
Furthermore, to each traffic class of each user in RRC\_INACTIVE\_CONNECTED, the subchannel for possible UL transmissions is allocated as a shared resource.
Here, traffic classes for each service are assumed to be registered at the initial service negotiation and RRC connection stage (i.e., admission).
If some traffic classes of a user require medium to low latency (for example, Class 2 in Table \ref{table2}), then the user can utilize the RRC\_INACTIVE state, in which the allocated preambles indicating the user identity and each of the traffic classes with such latency requirements enable a fast RRC connection setup once a packet of such classes arrives.
In addition, if some traffic classes of a user require very low latency (for example, Class 6 in Table \ref{table2}), the user can utilize the RRC\_INACTIVE\_CONNECTED state, where the allocated subchannel and user-specific preamble for each of such traffic classes enable immediate RRC connection resuming as soon as a packet of such a class arrives.

\begin{figure*}[t] 
	\centering
	\includegraphics[width=.8\textwidth]{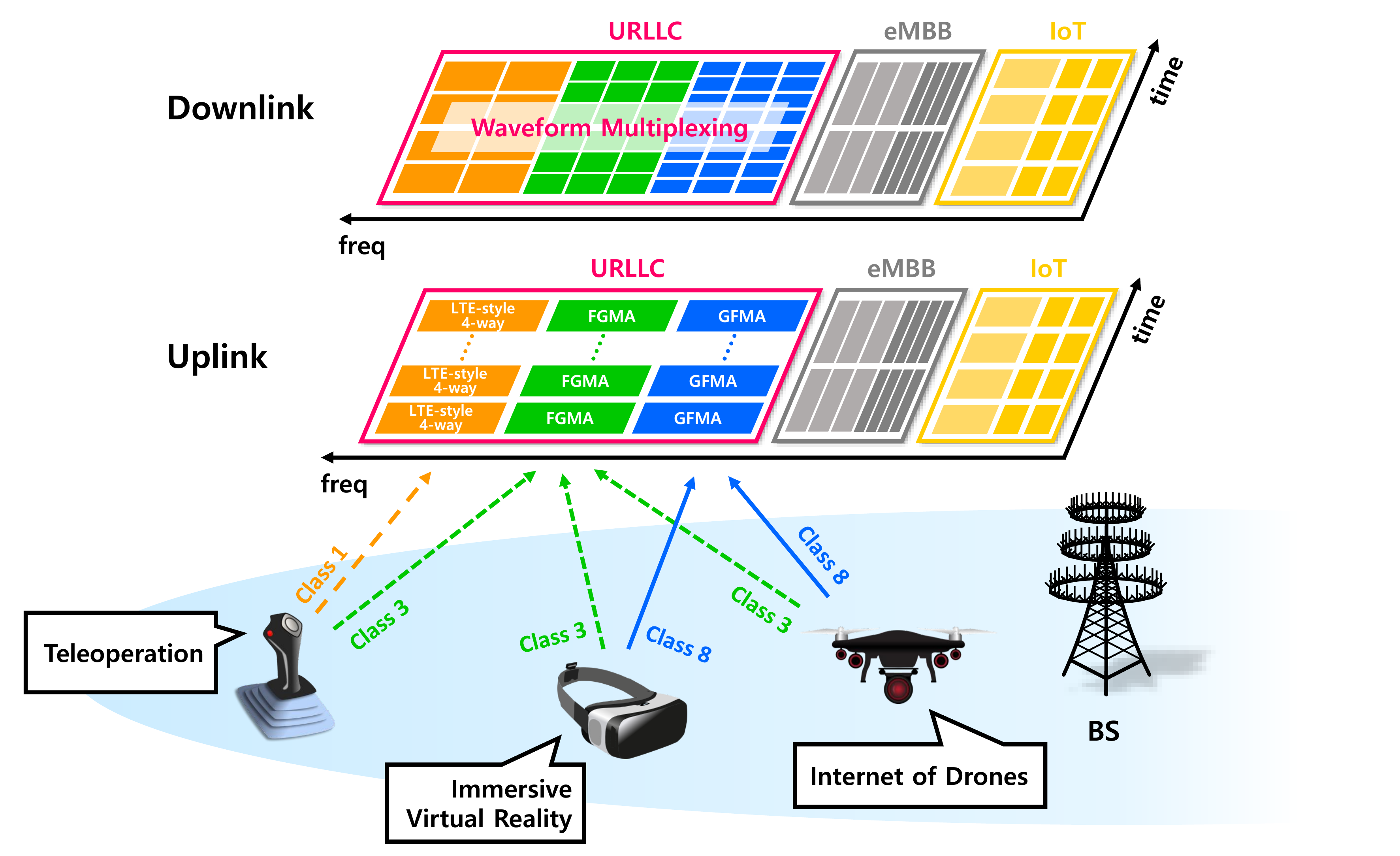} 
	\caption{Multiplexing of different multiple accesses for DL and UL.} 
	\label{fig-sysmodel}
\end{figure*}

The discussion above on the proposed RRC state transition can be re-drawn in a protocol perspective in Fig. \ref{fig-procedure}. Here, three protocols for the RRC connection are presented, in which the first one represents an LTE-style four-way handshaking RRC connection procedure, and the second one represents a two-way handshaking RRC connection procedure for providing a fast-grant multiple access (FGMA), and the last one represents an immediate RRC connection for providing a grant-free multiple access (GFMA). 

\begin{figure*}
	\centering
	\subfigure[Preamble indicating user and traffic class.]{%
		\includegraphics[width=.4\textwidth]{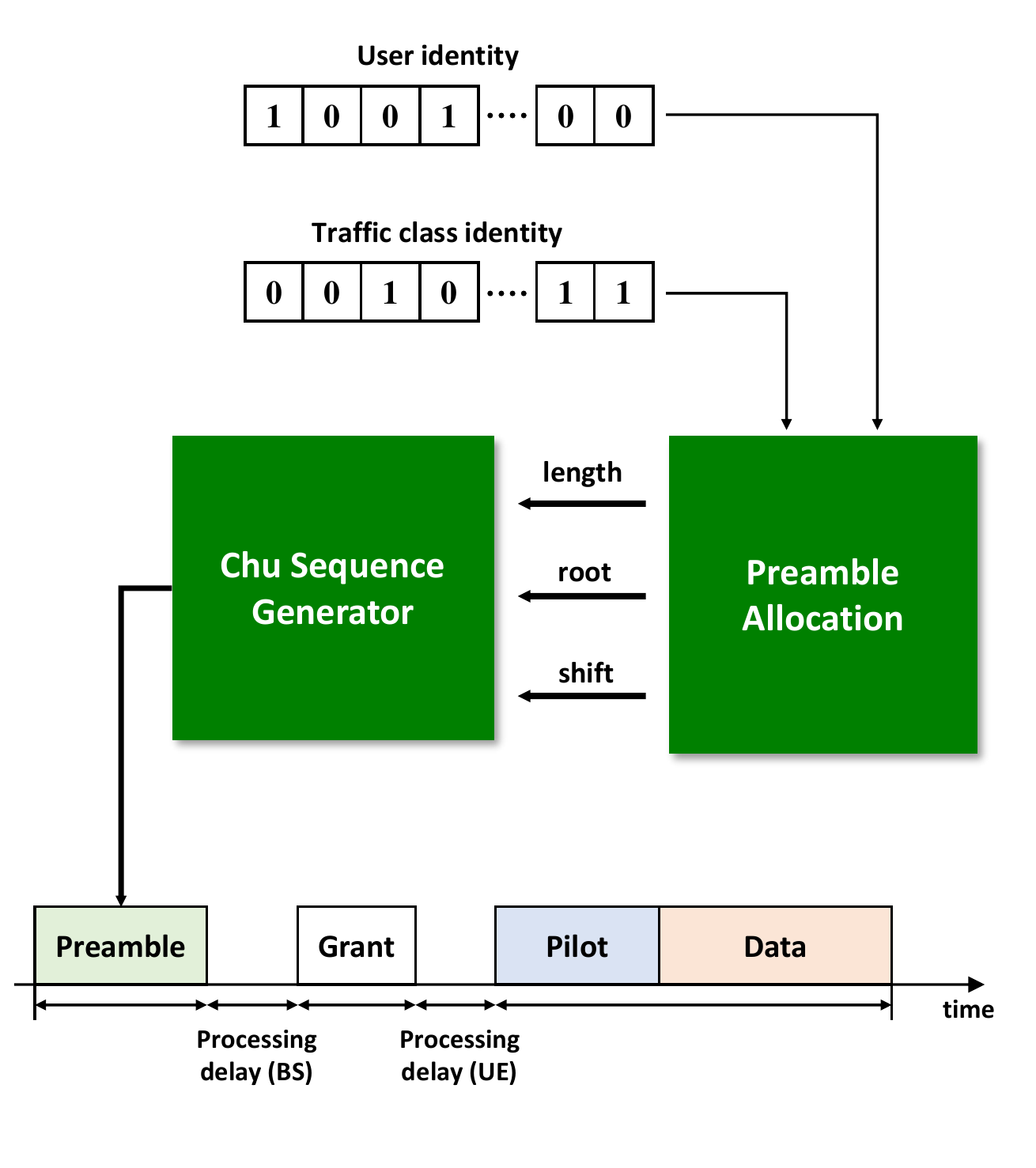}
		\label{preamble}
	} 
	\quad 
	\subfigure[Latency-optimal radio resource management and optimal frame configuration]{%
		\includegraphics[width=.5\textwidth]{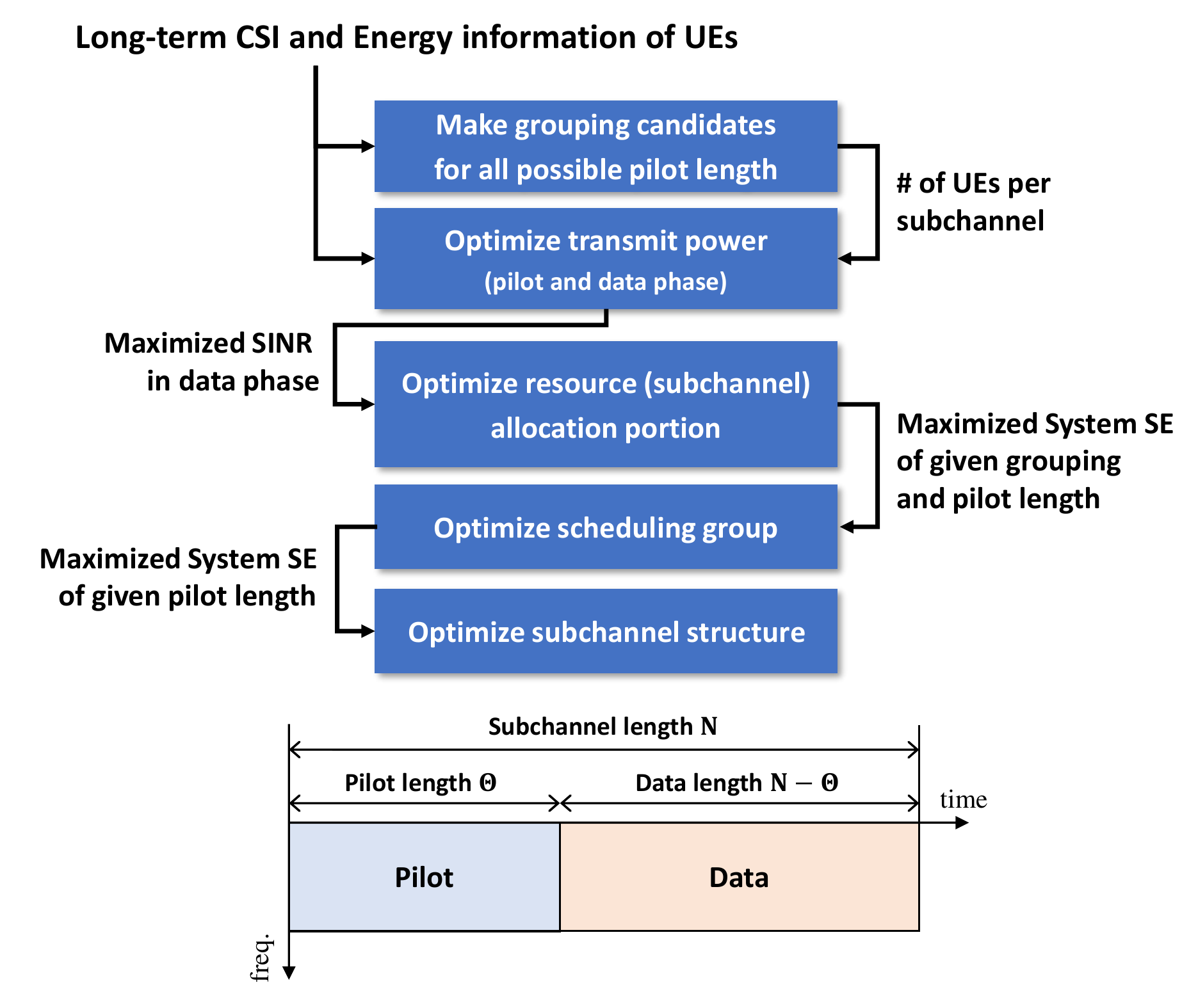}
		\label{alg}
	} 
	\caption{Optimal radio resource management strategy for FGMA satisfying latency and reliability requirements.} 
\end{figure*}

The first protocol is for traffic classes with loose latency requirements, such as in the first row in Table \ref{table2}. The LTE-style four-way handshaking using a cell-specific common set of preambles is spectrally-efficient and the LTE-style granted access is performed in the UL. However, as the latency becomes slightly tight, the delay caused by a grant procedure needs to be reduced and sending a scheduling request after a (sporadic or bursty) packet arrival is enough to obtain a grant for intended packets since a unique preamble indicating the user identity and the intended traffic class identity is already allocated and used in the scheduling request so that a base station performs a scheduling for the reliable delivery of such packets immediately and sends a grant with the allocated subchannel information. This protocol and FGMA is suitable for traffic with medium latency requirements as in the second row in Table \ref{table2} and traffic with low latency requirements but periodic arrival characteristics as in the third row in Table \ref{table2} so that a semi-persistent scheduling and subchannel allocation can be used.

For traffic with low latency requirements, the protocols above may not be used and an immediate packet transmission is required in the UL as soon as a packet of such a class arrives. In this case, the third protocol for GFMA is suggested, in which a subchannel as a shared resource and a user-specific preamble as a dedicated resource are already allocated for each traffic class with a low latency requirement and used for an immediate packet transmission as soon as a packet arrives. The GFMA with such a protocol is suitable for traffic with low latency requirements as in the fourth to eighth rows in Table \ref{table2}.

To employ such different multiple access schemes in a single carrier, the DL and UL radio resources are partitioned as shown in Fig. \ref{fig-sysmodel}.
For each service of each user, traffic is classified according to traffic characteristics and QoS as described in Section \ref{S2} and traffic of multiple users with similar characteristics and QoS are grouped and served together in each multiple access. Although different procedures for the RRC connection and the corresponding multiple access concepts are proposed to support various latency requirements required for URLLC services, providing reliability at a reasonably high spectral efficiency is still quite challenging. One good approach is to make the traffic requirements and QoS of multiple users in each FGMA or GFMA component as similar as possible and it can facilitate designing a radio resource management for reliability and high spectral efficiency \cite{ChoiSPS}.

\subsection{Multiple Access with Latency-Optimal Radio Resource Management}

Reliable information delivery in a cellular communication environment has been challenging because of channel quality fluctuation caused by the wireless fading channel and mobility. Although a low level of reliability could be provided by exploiting a limited order of diversity in time, frequency, and space in legacy cellular communications (the second generation or the earlier-stage third generation), the LTE has been successful in providing a high level of reliability primarily by using AMC based on channel quality information measurement and feedback and retransmissions using HARQ \cite{LTEbook}. 

However for URLLC services, a low latency requirement may restrict the use of AMC and HARQ or at least allow them only in a very limited manner so that most of the reliability part needs to be resorted on diversity again. Although classical repetition approaches can be adopted in time, frequency, or even multiple communication interfaces can be used \cite{Pop}, spectral efficiency may be significantly degraded, especially as more URLLC services are served. Thus, better approaches without significant spectral efficiency degradation are preferred and the most promising solution is to employ a large number of antennas at a base station, i.e., LSAS.  

Once an LSAS is assumed, the channel fluctuation caused by the wireless fading channel and mobility can be overcome (or significantly reduced at least) because of the channel hardening effect \cite{HochwaldChannelHardening}. However, the challenge is on the radio resource optimization in which preamble overhead, channel estimation quality, and user grouping are jointly considered and optimized.  
In \cite{ChoiSPS}, the authors proposed a latency-optimal semi-persistent scheduling algorithm for an LSAS, which can be utilized for guaranteed reliability in FGMA or GFMA. 

\begin{figure*}[t] 
	\centering
	\includegraphics[width=.8\textwidth]{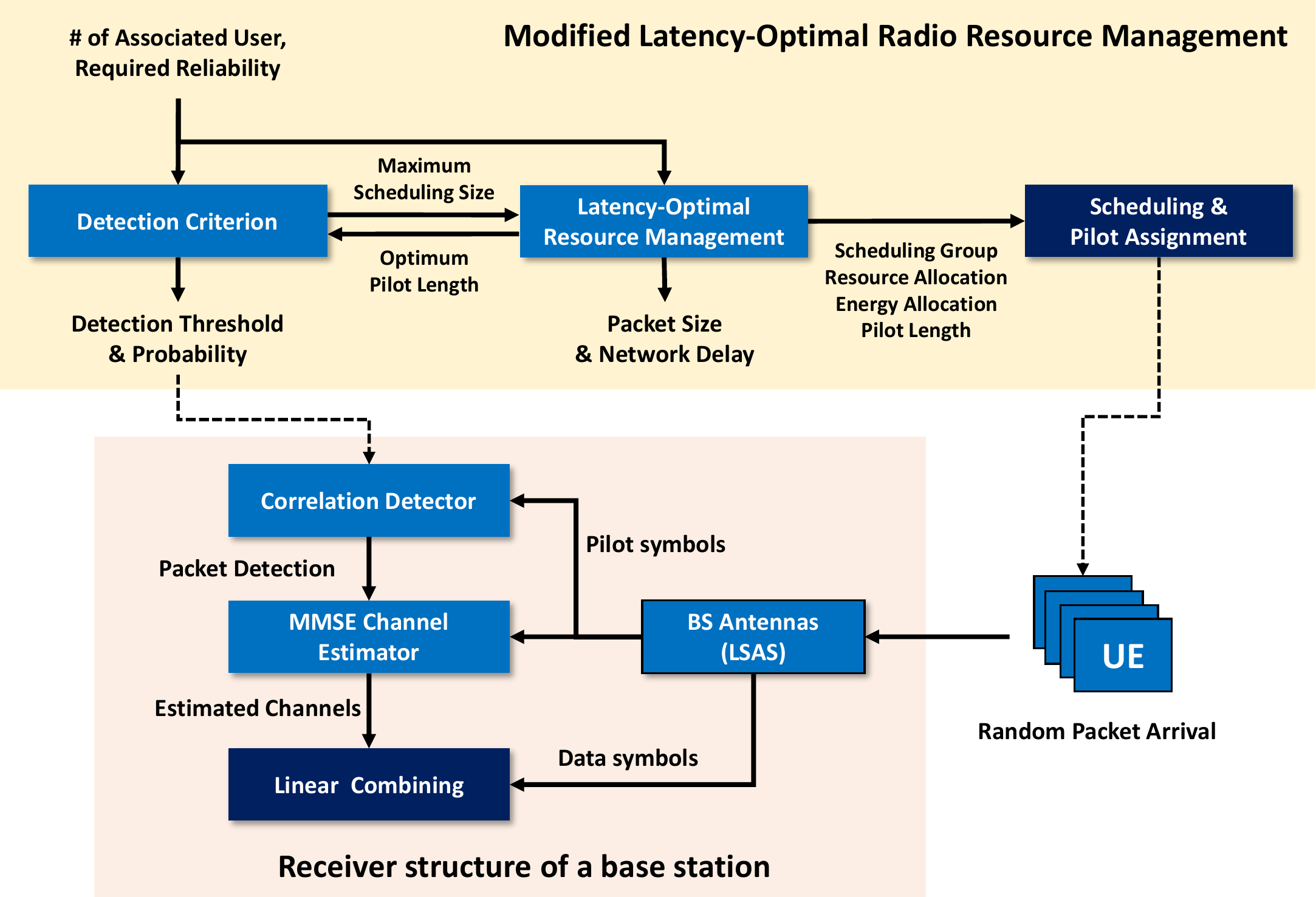} 
	\caption{Radio resource management concept and receiver structure for GFMA \cite{GFMAACCESS}.} 
	\label{GFMAaccess}
\end{figure*}

For FGMA, a unique preamble indicating the user identity and traffic class identity, as shown in Fig. \ref{preamble}, is allocated to each traffic class for each user during the admission control process. As the traffic characteristics and QoS of an arrived packet of each user, such as packet size, arrival model and rate, and latency and reliability requirements, can be detected at a base station from the preamble sent during its scheduling request, the base station can group users with similar traffic characteristics and QoS and apply the latency-optimal scheduling algorithm in \cite{ChoiSPS}. As shown in Fig. \ref{alg}, the transmit power of each user is first optimized based on the long-term channel state information and energy information of each user and then optimal user grouping is performed in which each user group shares a subchannel. From the optimization results, the pilot overhead for each subchannel is dynamically optimized as shown in Fig. \ref{alg} and the amount of resource to guarantee traffic delivery with reliability and latency requirements is determined. Subsequently, the subchannel construction and allocation information is delivered in a resource grant. Therefore, the proposed FGMA with a latency-optimal radio resource management can maximize the spectral efficiency while guaranteeing latency and reliability requirements for URLLC services. 

In GFMA, the challenge for guaranteed reliability is even more difficult because each user transmits its packet as soon as it arrives without any grant. However, employing an LSAS can also reduce such an uncertainty in addition to the channel hardening effect reducing the uncertainty in channel quality and it is possible to modify the algorithm in \cite{ChoiSPS} by considering such uncertainties together as indicated in Fig. \ref{GFMAaccess}.
{As traffic with similar characteristics and QoS is already grouped at the admission control stage,
a base station is aware of the arrival model and rate of each user so that it can be aware of the statistics for the actual transmitting users in each user group candidate.}
Thus, the base station can determine the optimal scheduling by considering such statistics of the user grouping candidates \cite{KimGFMA, GFMAACCESS}.
At the base station receiver, user detection needs to be first performed using preambles and the user detection capability should provide a success probability higher than the required reliability level, which is also considered in the optimization process \cite{KimGFMA, GFMAACCESS}. Simulation results in \cite{KimGFMA, GFMAACCESS} showed that the required radio resource for GFMA does not increase significantly compared with the case of a granted multiple access such that the proposed protocol and GFMA with a latency-optimal radio resource management can maximize the spectral efficiency while guaranteeing latency and reliability requirements for URLLC services. 

\section{More PHY technologies}
In the previous section, a set of multiple access techniques have been introduced, in which i) data packets for URLLC services are classified according to their traffic characteristics, including packet size and arrival statistics, and their latency and reliability requirements, ii) radio resources are partitioned to multiplex different multiple access components simultaneously, iii) each user or base-station is equipped with as many queues as its number of different packet classes, and iv) each multiple access supports its own packet class of multiple users. By virtue of the large number of antennas and the latency-optimal scheduling, the latency and reliability requirements of each packet class can be simultaneously satisfied.  

However, to realize such a concept, the radio resource needs to be well partitioned in a waveform level with good synchronization strategy. Moreover, to maximize the spectral efficiency, it is desired to use waveforms not only matched to user environment (i.e.,  delay spread and mobility) similar to the numerology multiplexing concept \cite{num}, but also appropriate for the latency requirements of users because the latency caused by the filters in a transceiver can be critical for packets with extremely low latency requirements. Thus, to devise a waveform multiplexing is a natural consequence, in which different types of waveforms (i.e., filtered-OFDM, generalized frequency division multiplexing (GFDM), etc) each with different numerologies (cyclic prefix length, subcarrier spacing, filter length, etc) are multiplexed. 

\begin{figure*}[t] 
	\centering
	\includegraphics[width=0.8\textwidth]{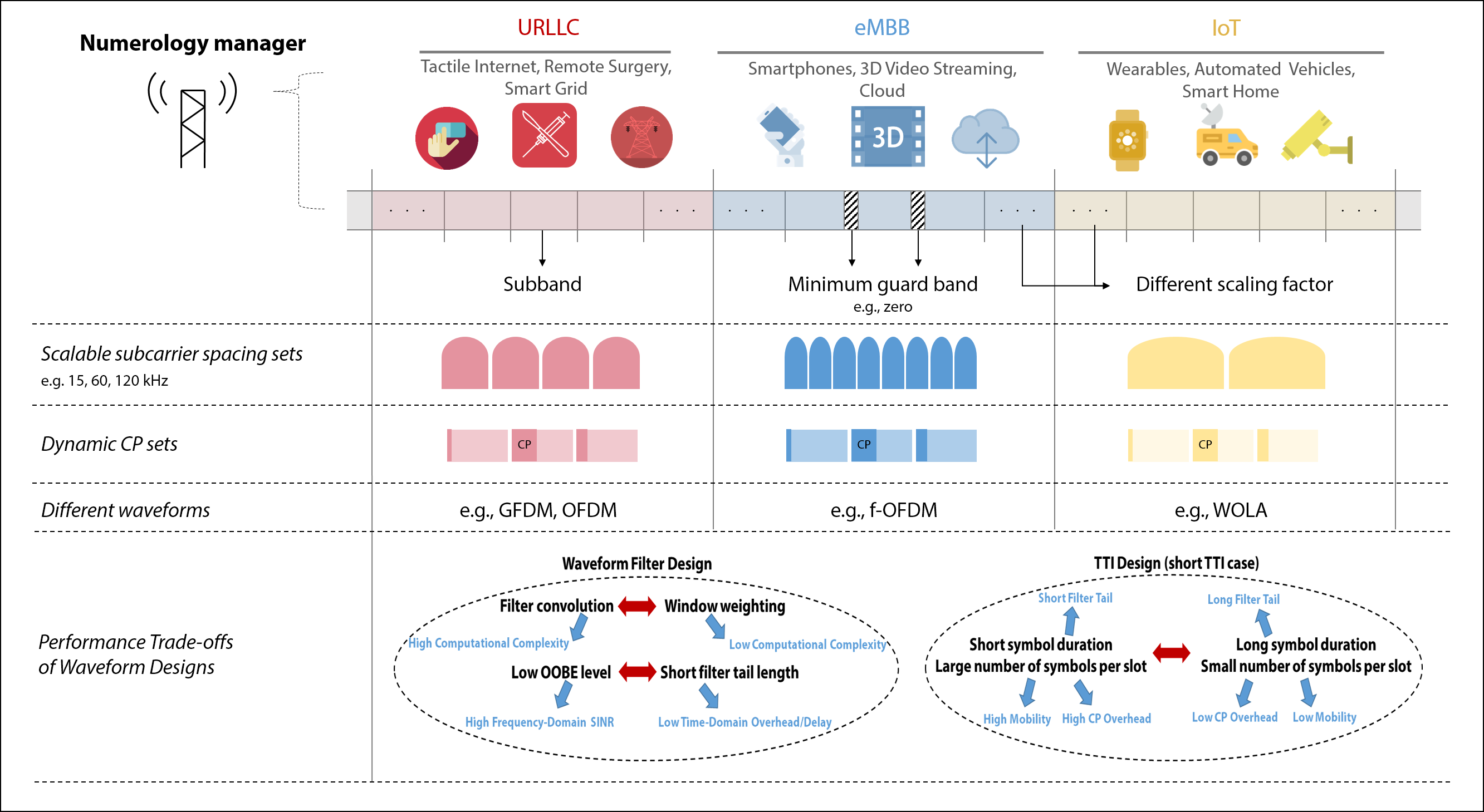}
	\caption{The proposed waveform multiplexing concept \cite{IEEE_WCM_WM}.} 
	\label{WM} 
\end{figure*}

In a latency budget such as in \cite{3gpplow}, one important component is the processing delay and most of the processing delay comes from the channel encoding/decoding latency. Thus, it is very important to devise channel codes with low encoding/decoding latency. In addition, such channel codes should have good performance in a high-reliability regime (i.e., frame error rate (FER) in the range of $10^{-3}$ to $10^{-7}$) such that both the water-fall performance and the error floor performance need to be considered.

To further improve the spectral efficiency and reduce the delay between the DL and UL, the best method is to adopt full duplex communication and the corresponding frame structure. Since an LSAS is assumed, a channel reciprocity such as in time division duplexing (TDD) is required for efficient channel estimations. However, in TDD, the delay between UL and DL subframes (or (mini) slots) may cause a latency problem. Thus, a practically feasible full duplex cellular communication technique can provide not only almost double the spectral efficiency but also a reduced delay between UL and DL as in frequency division duplexing (FDD). Although the feasibility for self interference cancellation (SIC) at a (low-power) base station has been confirmed \cite{fulld,Chung2017}, the interference at DL users caused from UL users needs to be avoided. Although a full-duplex cellular communication can work if an appropriate paring of DL/UL users is assumed \cite{fully}, better interference avoidance scheme needs to be devised for incorporation with various scheduling strategies without such pairings.
 
\subsection{Waveform Multiplexing}

To multiplex different classes of services in a common carrier, one approach is to multiplex URLLC packets on eMBB resources, such as in \cite{DLmux2}, and the other is a numerology multiplexing by resource partitioning, such as in \cite{DLmux1}, where different numerologies for OFDM parameters and frame structure can coexist. Owing to the capability of selecting the appropriate numerologies according to the users' environments and service requirements, a numerology multiplexing is considered as a promising solution. However, inter-numerology interference needs to be taken into account and an appropriate filter design for low out-of-band-emission (OOBE) is required \cite{IEEE_ComMag_WF}.

As an elegant solution to implement such a numerology multiplexing concept, a waveform multiplexing system is proposed in \cite{IEEE_WCM_WM} as illustrated in Fig. \ref{WM} and it employs scalable subcarrier spacings and dynamic cyclic prefic (CP) managements. The proposed waveform multiplexing selects not only appropriate subcarrier spacings and CP lengths according to users' channel environment and mobility but also waveform filters for minimum guard bands according to service (latency) requirements and OOBE levels. 

In the proposed waveform multiplexing, users with similar mobility (channel coherence time), delay spread, and latency requirements are grouped and the appropriate subcarrier spacing and CP length are selected for each group to minimize the CP overhead. For each group, an appropriate waveform filter is determined to minimize the guard band while satisfying latency requirements. In cases where a low OOBE level is desired and latency requirement is loose, waveforms with very low OOBE, such as in \cite{f-OFDMmux,f-OFDM1,f-OFDM2}, can be used to enhance the frequency-domain SINR. However, in cases where an extremely low latency is required, waveforms with short filter delays at a reasonable OOBE, such as in \cite{GFDM1,GFDM2}, may be preferred. Such a waveform multiplexing concept can be considered a genealization of numerology multiplexing in a single waveform, such as in \cite{f-OFDMmux,FC-OFDM}. 

\begin{figure*}
	\centering
	\subfigure[The proposed receiver filter structure at a base station.]{%
		\includegraphics[width=.45\textwidth]{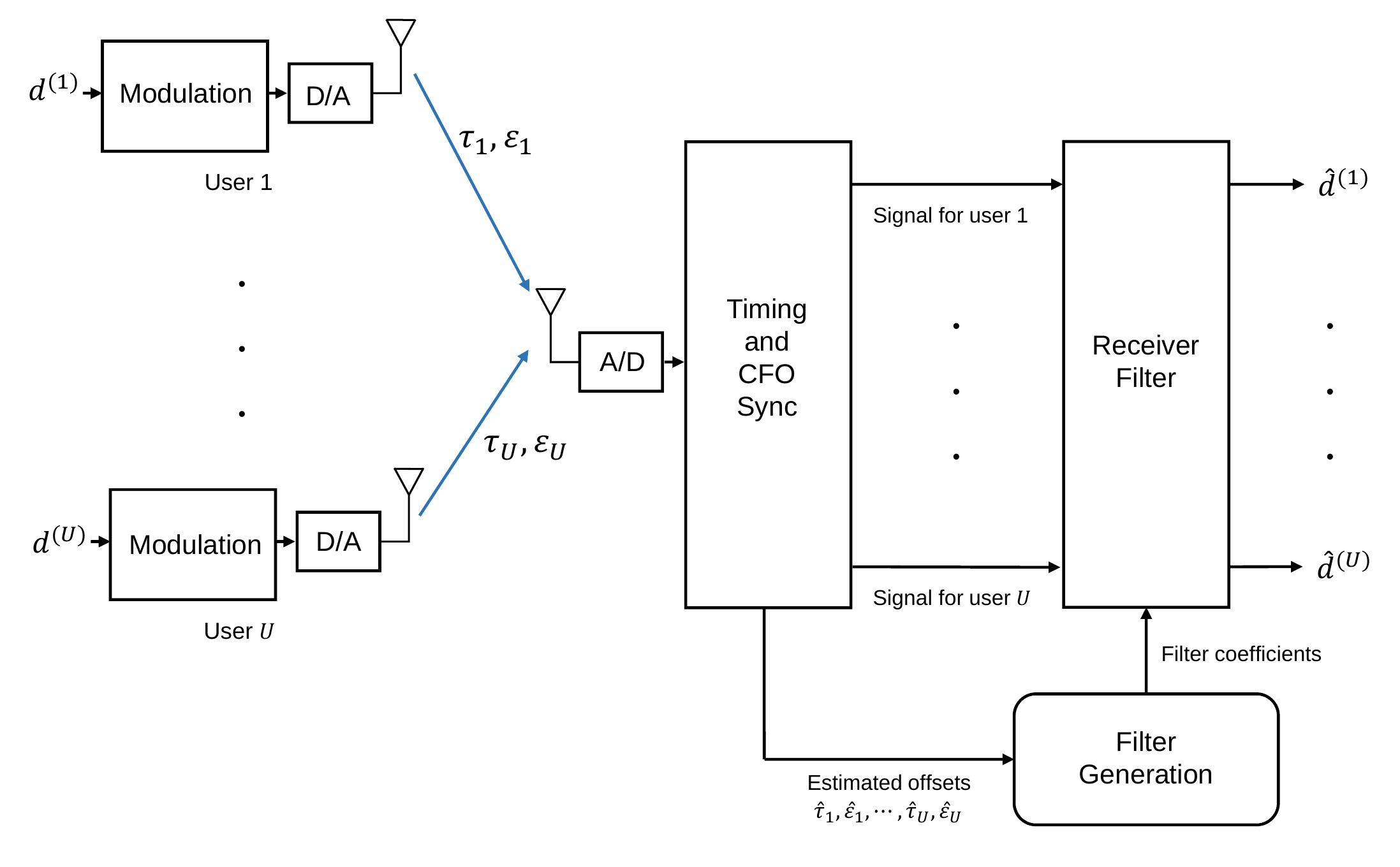}
		\label{sync}
	} 
	\quad 
	\subfigure[Performance comparison.]{%
		\includegraphics[width=.4\textwidth]{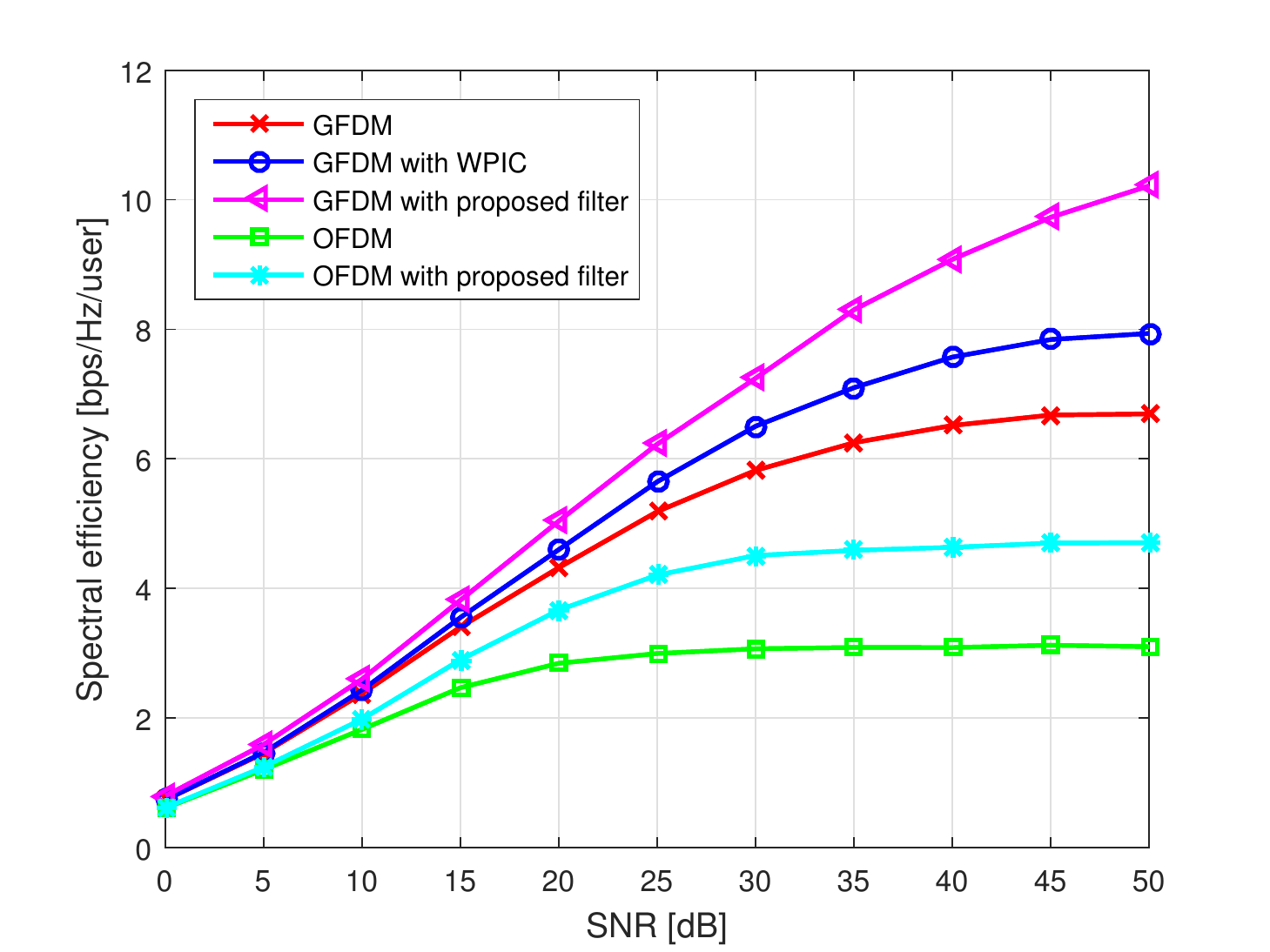}
		\label{sync2}
	} 
	\caption{UL receiver structure for handling synchronization issue \cite{sync_twc,sync_wcnc}.} 
\end{figure*}

\begin{figure*}
	\centering
	\subfigure[Protograph structure.]{%
		\includegraphics[width=.35\textwidth]{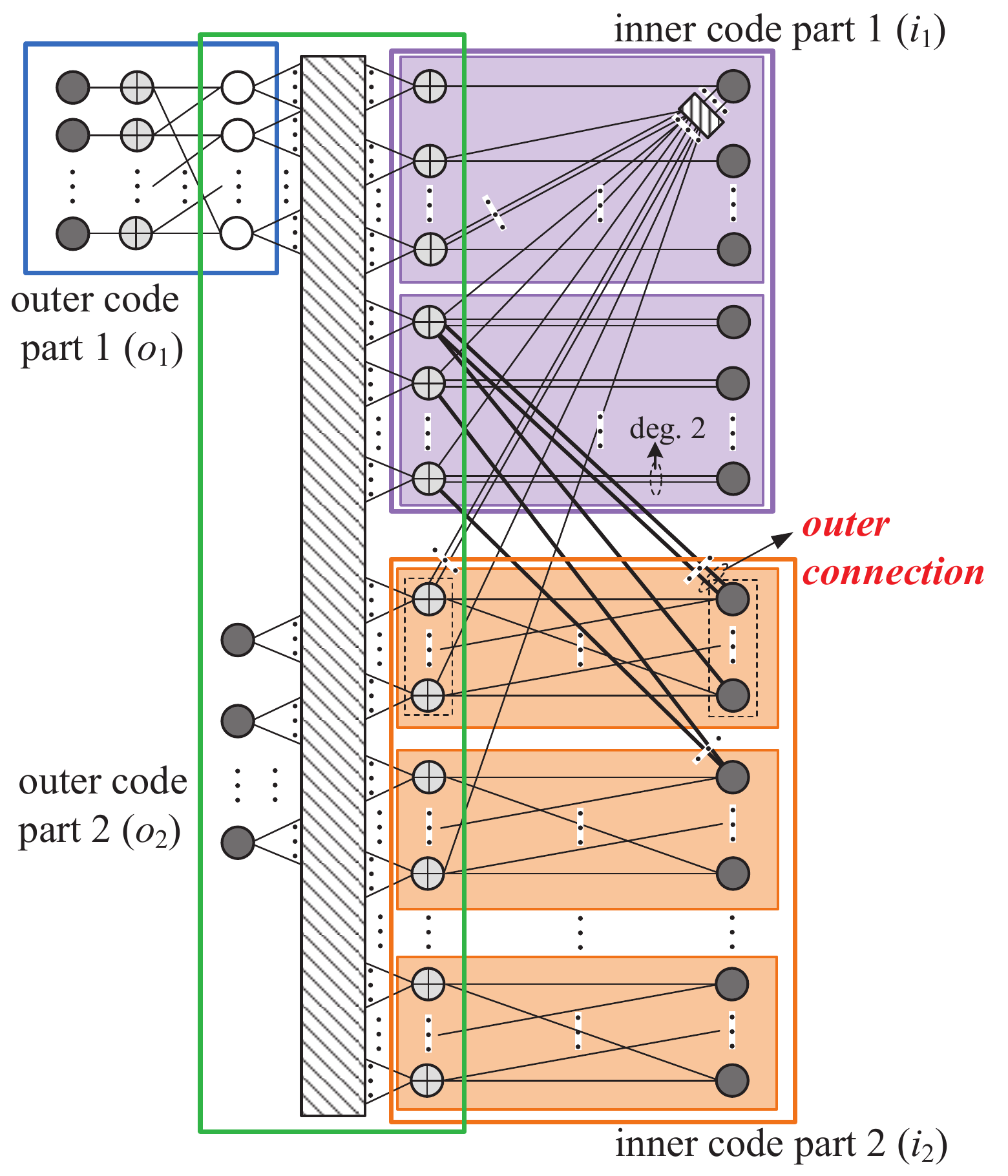}
		\label{ARACA1}
	} 
	\quad 
	\subfigure[Required SNRs for target FERs.]{%
		\includegraphics[width=.5\textwidth]{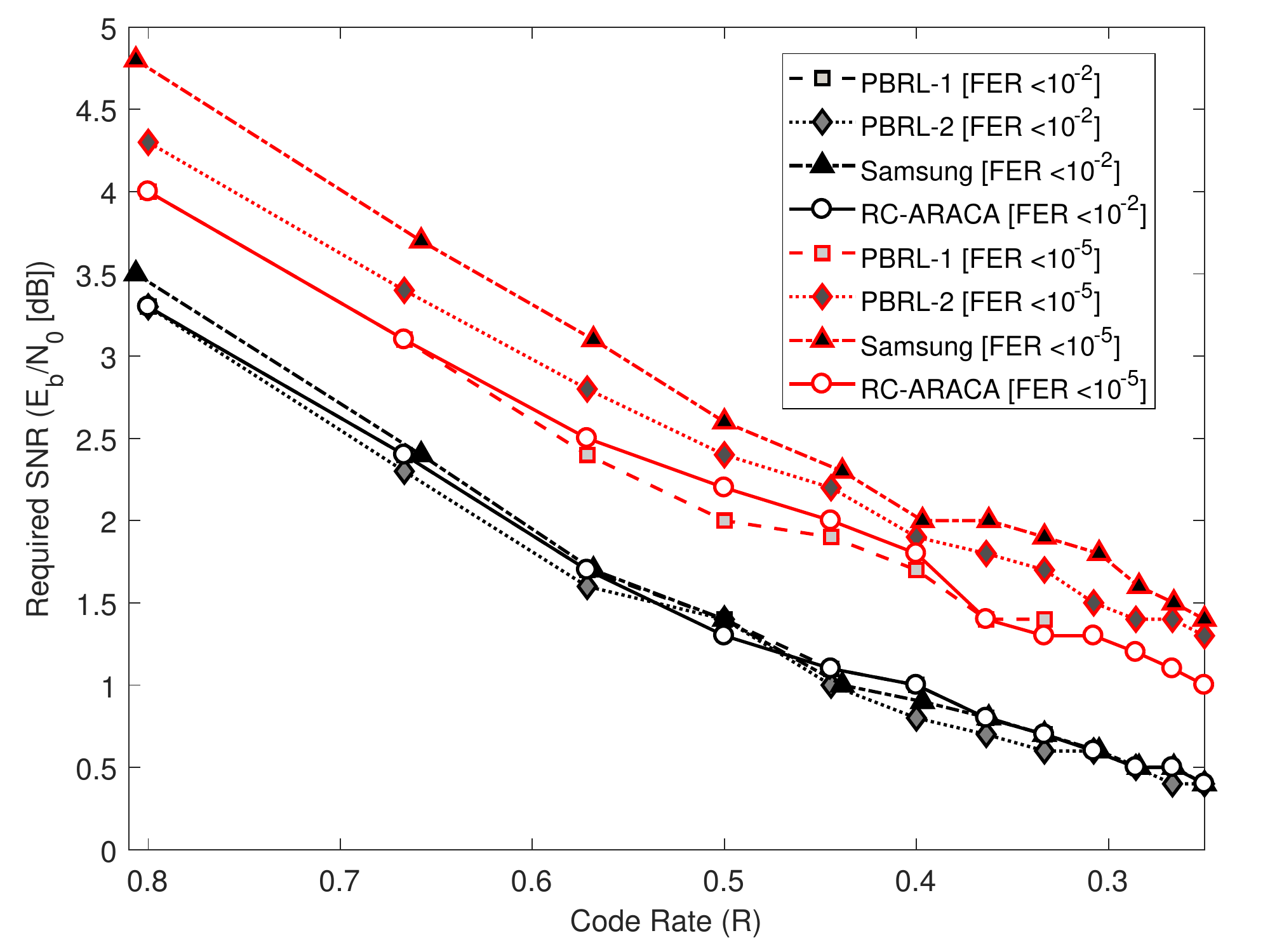}
		\label{ARACA2}
	} 
	\caption{Protograph structure of an ARACA code and performance comparison \cite{ARACA}\cite{RCARACA}.} 
\end{figure*}

\subsection {Synchronization Issue}
In the DL, each user can employ a legacy time and frequency synchronization, such as in \cite{Morelli,Gaspar}, on the subbands where it belongs to, even in cases of employing a waveform multiplexing with different numerologies and waveform filters. Thus, synchronization for DL does not raise a new critical issue and can be done similarly as in the LTE. 

In the UL, it is reasonable to assume a similar closed-loop procedure for a strict time synchronization as in the LTE for eMBB and URLLC services. However, since higher mobility and higher frequency bands need to be supported, especially for URLLC services, time synchronization errors and frequency offsets due to Doppler shifts may cause non-negligible performance degradation, especially for URLLC services in which high reliability is required and sporadic access needs to be supported. As a remedy, an interference cancellation approach is adopted at the base-station receiver \cite{sync_twc,sync_wcnc} as illustrated in Fig. \ref{sync}. Here, different time and frequency offsets of multiple users are assumed to be estimated similarly as in \cite{Morelliuplink,Beek} along with a closed-loop time synchronization as in the LTE. In order to reduce multiple user interference caused from time and frequency offsets of multiple users which may be caused from high mobility, high frequency, or sporadic access, an elaborately designed receiver filter is applied to maximize the signal-to-interference ratio (SIR) of users by using the estimated time and frequency offsets and it is shown in \cite{sync_twc,sync_wcnc} that the proposed approach can provide better performance compared to those in \cite{Manohar,Huang} as shown in Fig. \ref{sync2}.

\begin{figure*}
	\centering
	\subfigure[CDD-SDMA concept.]{%
		\includegraphics[width=.35\textwidth]{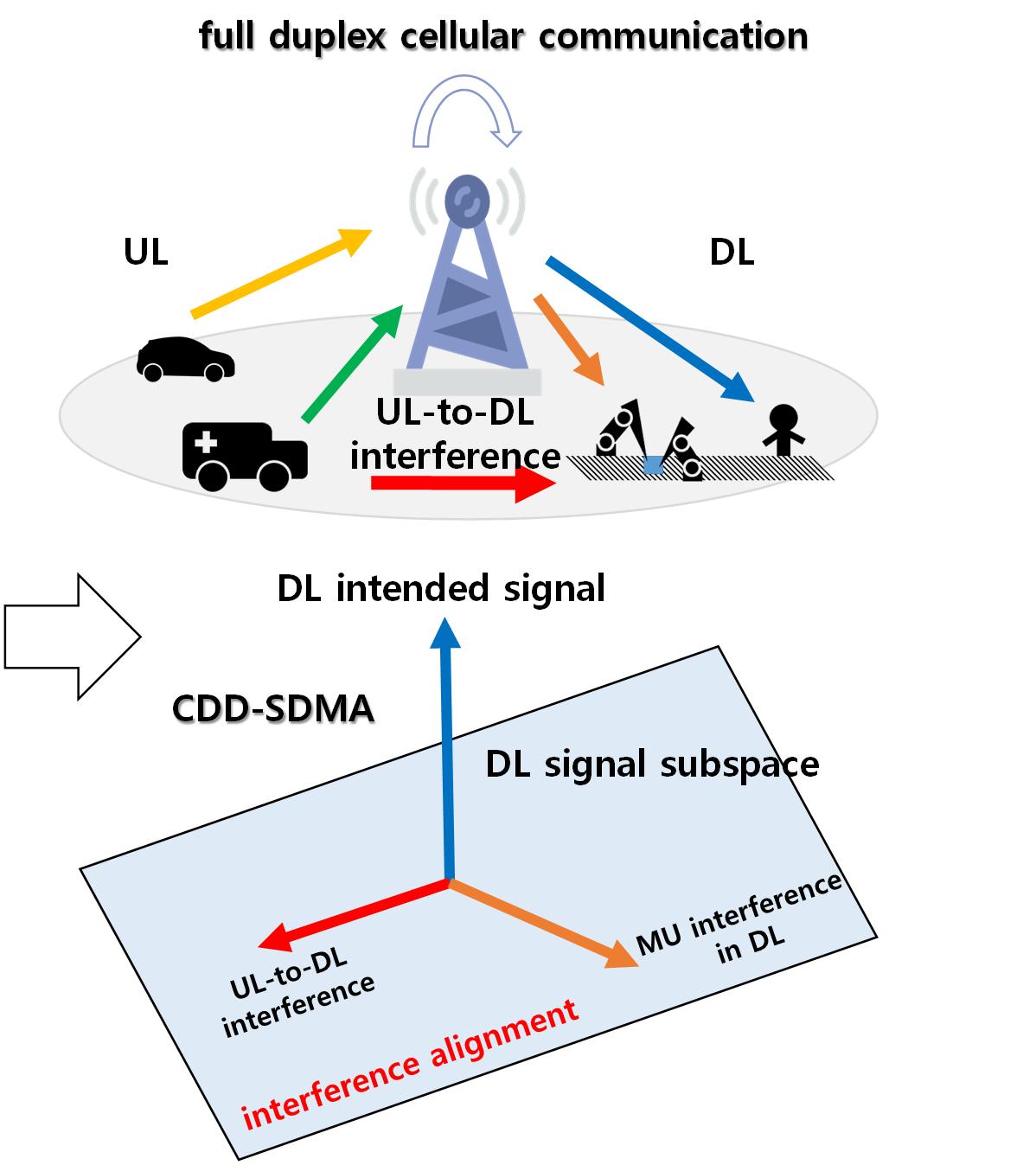}
		\label{cdd1}
	} 
	\quad 
	\subfigure[Performance evaluation.]{%
		\includegraphics[width=.45\textwidth]{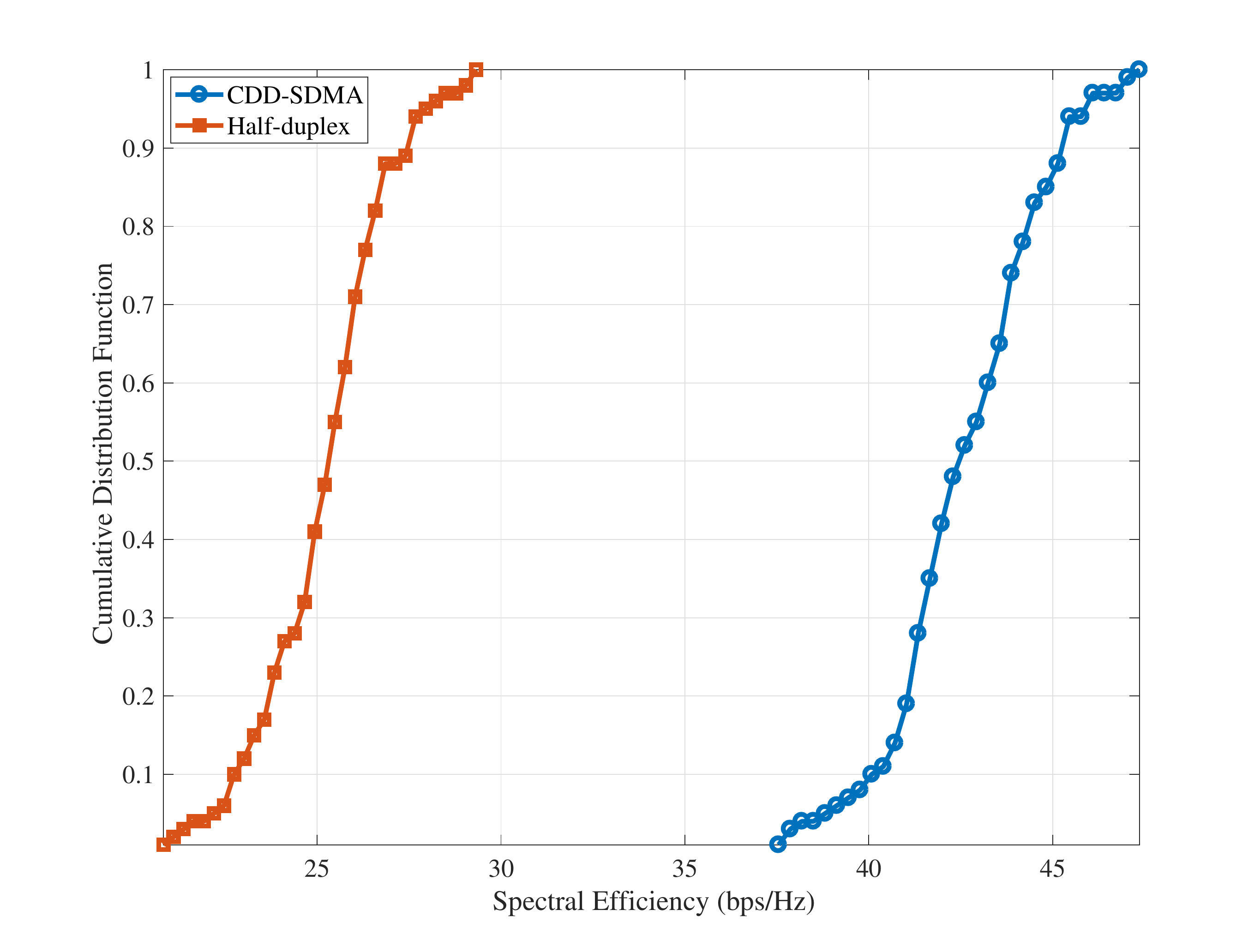}
		\label{cdd2}
	} 
	\caption{CDD-SDMA concept for a full duplex cellular communication and performance evaluation.} 
	\label{CDD}
\end{figure*}

\subsection{Channel codes for URLLC}

Recently, low-density parity check (LDPC) codes have been adopted for eMBB services in the NR standard \cite{5Gcc}. Such LDPC codes can be considered as raptor-like quasi-cyclic LDPC (QC-LDPC)  codes and they can provide near-optimal water-fall performance as well as efficient encoding and decoding implementation methods such that they are quite appropriate for eMBB applications.
However, their error floor performance may not be good, especially as the code rate decreases, because of the lack of linear minimum distance growth (LMDG) property and too many degree-1 variable nodes, as expected in  \cite{Error}, and it may limit the use of such protograph-based raptor-like (PBRL) QC-LDPC codes for URLLC applications, especially for the cases where the required reliability is quite high (e.g., FER in the range of $10^{-3}$ to $10^{-7}$) and the latency requirement is tight such that a retransmission is not allowed.

In \cite{ARACA}, accumulate repeat accumulate check accumulate (ARACA) codes are recently proposed by the authors to provide high reliability by having both the LMDG property (i.e., no error floor) and good water-fall performance with an efficient encoding structure. Fig. \ref{ARACA1} shows the protograph structure of an ARACA code, which is comprised of the two outer code parts (o1 and o2) and the two inner code parts (i1 and i2) as described in \cite{ARACA}, and it is characterized by the outer connections that can provide an efficient low-complexity encoding similar to an accumulate repeat accumulate code \cite{ARA} as well as the LMDG property with a water-fall performance similar to an accumulate repeat jagged accumulate code \cite{ARJA}. Further, Fig. \ref{ARACA2} shows the good performance of rate-compatible ARACA codes \cite{RCARACA} compared with PBRL QC-LDPC codes in \cite{SSS,RCSL,PBRL}. In addition, \cite{ARACA2} proposes a low-latency and low-complexity layered Richardson-Urbanke encoding method and encoder structure as well as a low-latency and low-complexity big-layer parallel decoding method and decoder structure, which shows that the proposed ARACA codes are promising for URLLC services.

\subsection{Full-duplex Cellular Communication: code-division duplexing spatial-division multiple access (CDD-SDMA)}

In a full-duplex cellular communication, DL users suffer from the interference caused by UL users as illustrated in Fig. \ref{cdd1}. As a result, without an elaborate management of such interference, the overall performance, such as the rate distribution of users, cannot be meaningfully improved, even in cases where perfect SIC is assumed at a base station.

As a remedy, a novel CDD-SDMA is proposed, in which UL interferences are aligned to a null space orthogonal to the signal subspace for DL multiuser multiple-input multiple output (MU-MIMO) by using orthogonal codes between DL and UL and employing antenna reconfiguration (or different versions of analog beamforming) to align all UL interferences into a single dimension of the DL signal space similarly as in \cite{Cadambe:08} and devising an efficient DL/UL MU-MIMO schemes on the remaining signal subspaces on DL/UL similarly as in \cite{Yang:17} with the DoFs approaching that of the normal zero-forcing MU-MIMO in a half-duplex DL/UL. 

To confirm the feasibility of the proposed CDD-SDMA, not only a practical indoor hotspot environment and a spatial channel model in \cite{3DSCM} are used but also adjacent channel interference and in-band blocking due to the remaining frequency offsets among UL users ($<100$Hz) and the finite resolution (12 bits) of analog-to-digital converters as well as co-channel interferences at the same resource block are considered in the case of 40 baseband streams and 100 physical antennas at a base station. As shown in Fig. \ref{cdd2}, more than $70\%$ improvement in spectral efficiency is expected in a practical environment, even considering such non-ideal effects. Thus, the proposed CDD-SDMA can be considered as a promising solution, not only for almost doubled spectral efficiency but also to significantly reduce the delay between UL and DL while exploiting channel reciprocity.

\begin{figure*}[t] 
	\centering
	\includegraphics[width=\textwidth]{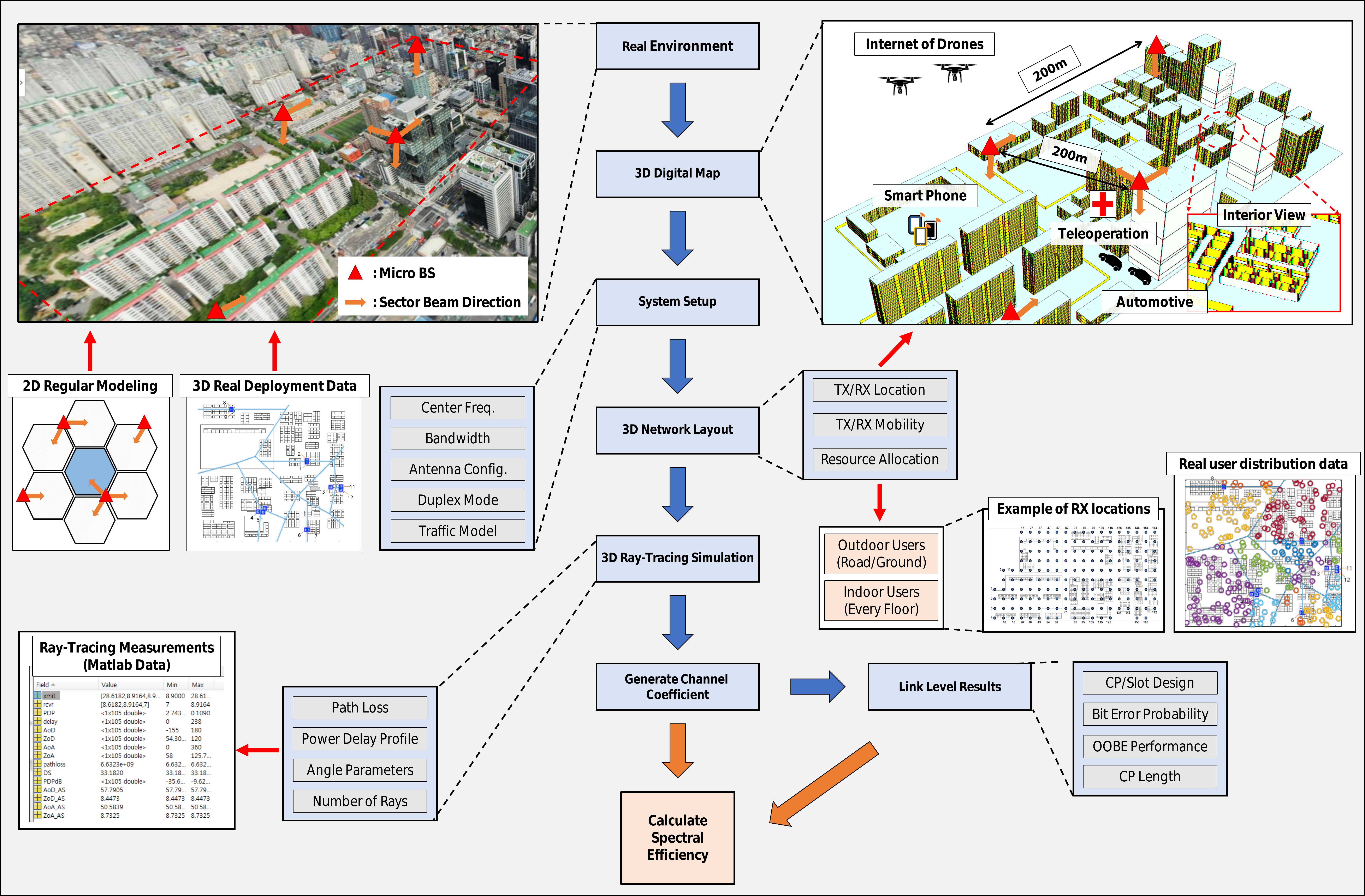}
	\caption{Proposed evaluation methodology.} 
	\label{SLS} 
\end{figure*}

\section{Evaluation Methodology and Simulation Results}

Two-dimensional (2D) regular cell layouts with 2D stochastic wireless fading channel models have been widely used to evaluate the performance of legacy cellular systems. As the uses of multiple antennas and small cells become widespread,  beam-steering effects according to elevation angles matter and such 2D channel models with a 2D regular cell layouts have evolved into 3D channel models by applying stochastic channel parameters for elevation angles, which include the 3GPP 3D spatial channel model~\cite{3DSCM} used for evaluating LTE technologies. However, in most LTE system-level simulation (SLS) scenarios, 2D regular layouts have been commonly used such that the channel parameters between a randomly selected transmitter and receiver pair are primarily dependent on scenario-dependent parameters and the locations of nodes including the two in the desired pair and interfering sources. 

As a typical cell size shrinks for an enlarged area spectral efficiency, a cell deployment scenario should consider the geography of a target environment including its landform, shapes and heights of surrounding structures such as buildings, and different attenuation factors due to different constituent materials of each surrounding structure. To exploit such a real geography, map-based channel models utilizing ray-tracing tools have drawn much interest from academia and industry, such as in \cite{MapCh_Rapa, Wise_Mag, MapCh_Heath,MapCh_3GPP,Map_METIS,Map_METIS_Mag,Map_Lim}. Reasonable agreement with hardware measurements has been reported in \cite{MapCh_Rapa,Wise_Mag,MapCh_Heath,Map_METIS,Map_METIS_Mag,Jang_Smallcell} and link-level simulations (LLSs) and SLSs have been performed to evaluate their proposed work by using measurements from hardware testbeds and/or software algorithms in environments similar to real worlds \cite{IEEE_WCM_WM,Oh_DP,Jang_Smallcell,Kwon_TMTT_Lens,Sim_FD,JSAC_mmWaveRT,Access_mmWaveRT,Kim_NOMA,Lim_DP,Cho2018}.

Fig. \ref{SLS} shows the proposed 3D SLS evaluation strategy in this paper for the performance of UL multiple access and DL waveform multiplexing, where a high-resolution digital map is constructed for the GangNam station area (Seoul, Korea), real base station deployment information, such as the locations as well as the antenna heights and tilting angles, are taken into account, and the reported typical user density for each part of the digital map 
is applied as in \cite{IEEE_WCM_WM,KICS_RYU}. Using such a realistic digital map and the locations for base stations and users, 3D channel parameters are collected \cite{IEEE_WCM_WM, Map_Lim, KICS_RYU} using the ray-tracing tool called Wireless System Engineering (WiSE) developed by Bell Laboratories \cite{Wise_Mag}. Subsequently, based on the collected data, 3D wireless channels are generated as in \cite{3DSCM} according to either a deterministic model based on specific locations of the transmitter and receiver pairs or a stochastic model with statistics matched to this specific environment according to this digital map.  

\begin{figure*}
\centering
  \subfigure[Latency distribution  of URLLC users.]{%
    \includegraphics[width=.45\textwidth]{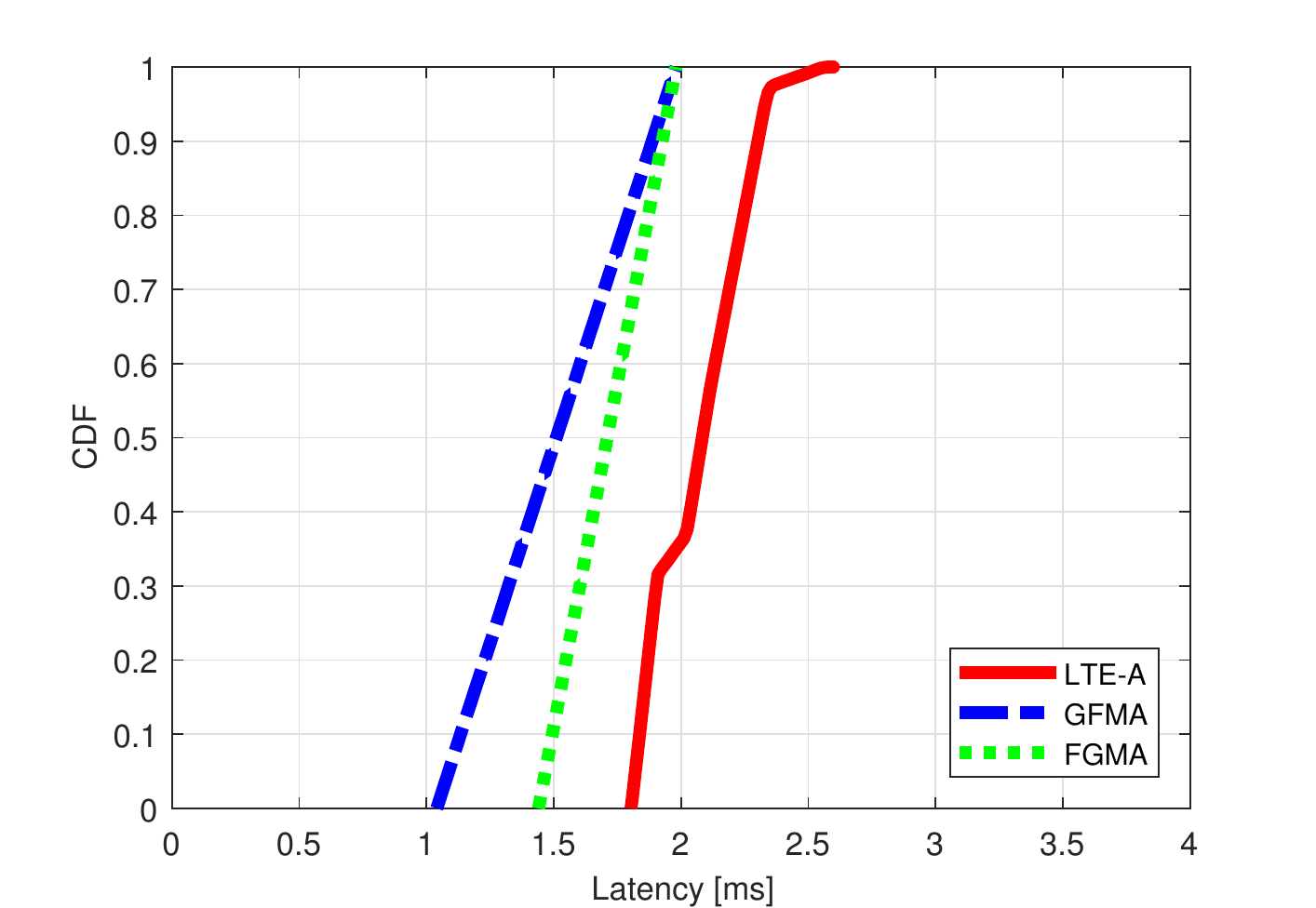}
    \label{latency}
  } 
  \quad 
  \subfigure[Spectral efficiency distribution for URLLC resource.]{%
    \includegraphics[width=.45\textwidth]{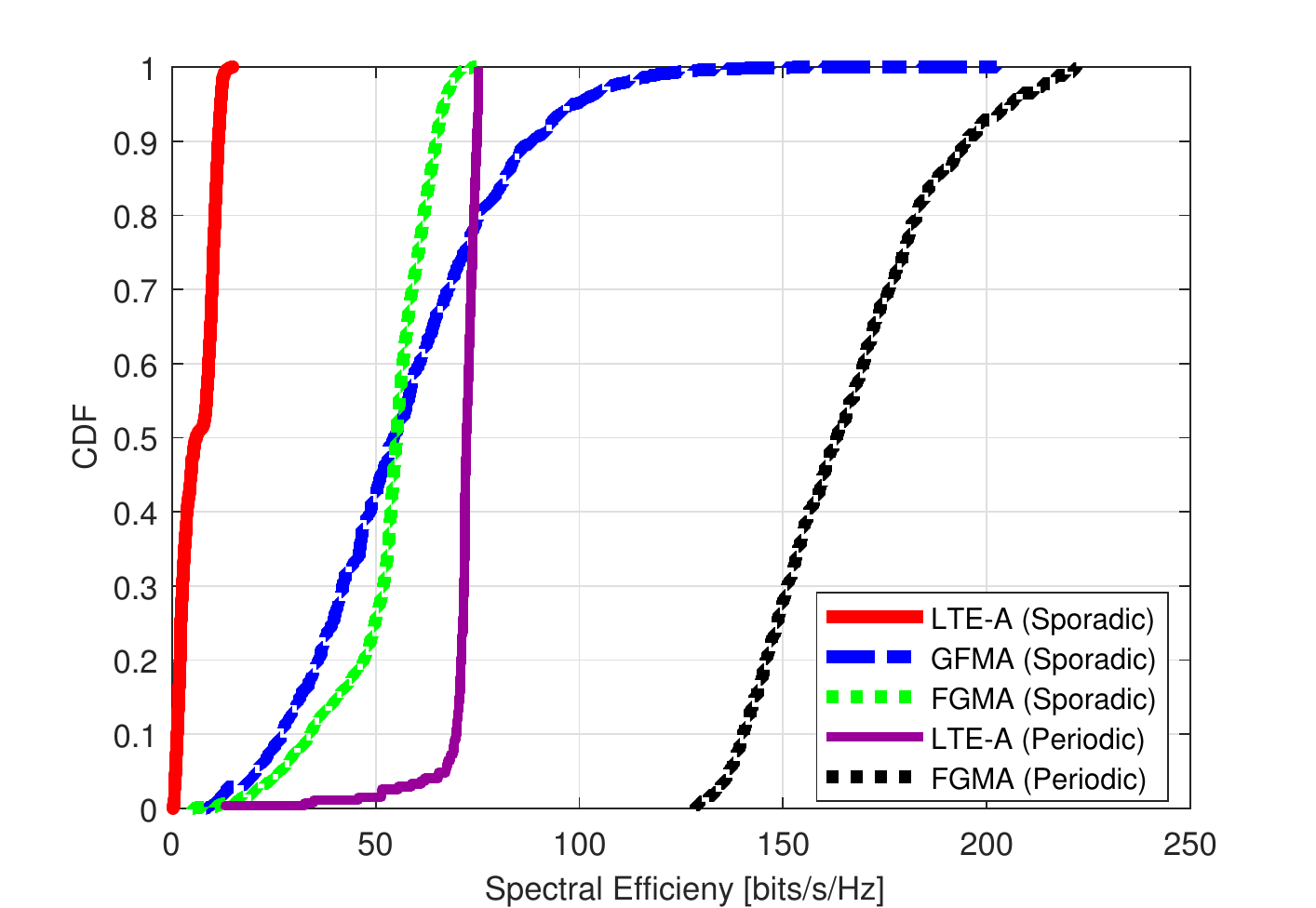}
	\label{SE}
  } 
  \caption{UL Performance evaluation using GFMA for a URLLC service.}  \label{GFMAPER}
\end{figure*}

\begin{figure*}[t] 
	\centering
	\includegraphics[width=.55\textwidth]{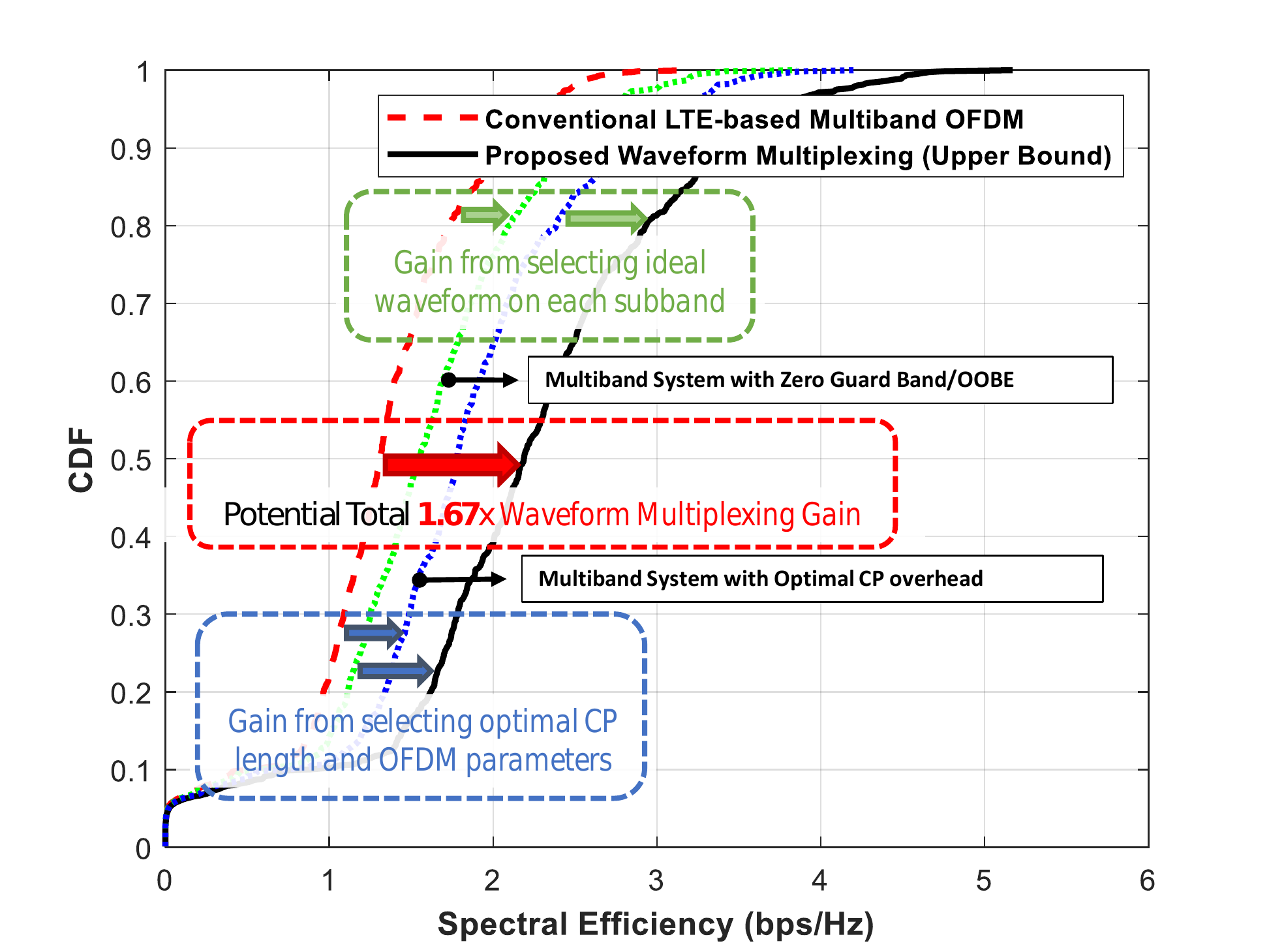}
	\caption{Achievable waveform multiplexing gain} 
	\label{WMgain} 
\end{figure*}

In Fig. \ref{GFMAPER}, the UL latency and spectral efficiency distribution of typically distributed 6000 RRC\_INACTIVE\_CONNECTED users are shown when an antenna array with 128 antenna elements is assumed in each of 12 real base stations deployed in the GangNam station area with the same height and tilting angle. Here, the traffic characteristics are assumed as follows: the packet size is 8 KB (64 Kbits), average arrival rate is 100 packets per second, arrival model is a sporadic Poisson random arrival and the latency and reliability constraints are 2 ms and 99.999\%, respectively. Here, the CP overhead is assumed to be 25\% and variable-length (minimum 0.2 ms) mini-slots comprised of subchannels are assumed for a frame structure. Further, users are associated with the nearest base station and each base station determines the required amount of radio resources (mini-slot length and bandwidth) to support grant-free accesses with guaranteed QoS for its associated users, in which user grouping, portion of pilot symbols in each subchannel allocated to each user group, and power control for each user are optimized. Then, the spectral efficiency is evaluated as the ratio of the sum of goodputs (in bps) and the sum of required amount of bandwidth (in Hz) of these cells considering the CP/pilot overhead, channel estimation error, and channel code FER performance. In addition, to evaluate the latency distribution, the required TTI length as well as queuing delay due to a random packet arrival, wireless propagation delay, and decoder processing delay are also taken into account, similarly as in \cite{3gpplow}, where the queuing delay is assumed to be uniformly distributed over a minimum TTI length and the throughput and latency of the channel decoder at a base station are assumed to be 50 Gbps and equal to the minimum TTI length, respectively. 

In Fig. \ref{latency}, the latency distribution of the proposed GFMA is evaluated and compared with the cases where an LTE-style four-way access with a round-robin scheduling and equal power control (denoted as `LTE-A Extension') and the proposed FGMA are instead applied. Here, in addition to the processing delay for decoding, a processing delay for scheduling as long as two times the minimum TTI length is considered in FGMA. 
Further in Fig. \ref{SE}, the spectral efficiency distribution (for goodput only) of the proposed GFMA is evaluated and compared with the two cases. From the results, it is confirmed that the proposed GFMA is the most efficient for traffic with tight latency requirements and sporadic arrival characteristics. In `LTE-Extension', a large amount of latency budget (more than 80\%) is wasted for the 4-way handshaking such that the latency distribution and spectral efficiency distribution for goodput are significantly degraded. In FGMA, although some portion in the latency budget is spent for the two-way handshaking and scheduling, a granted access with a latency-optimal scheduling improves the spectral efficiency during data mini-slots so that these two schemes can provide similar spectral efficiency performance in this specific case. In general, as the latency requirement becomes tighter and/or more antennas are equipped at a base station, GFMA performs better than FGMA.  
Also in Fig. \ref{SE}, the spectral efficiency distribution of FGMA is shown when the packet arrival model changes to be periodic with perfectly aligned arrival times so that a semi-persistent scheduling and resource allocation can be allowed. In this case, FGMA can provide much higher spectral efficiency with guaranteed latency and reliability because the fast protocol of FGMA enables an initial access with guaranteed latency requirement and such initial overhead for the grant becomes ignorable.

In summary, Fig. \ref{GFMAPER} shows that i) although equipped with a large number of antennas and reduced TTIs, LTE-style RRC connection protocol and multiple access cannot provide sufficiently high reliability and low latency even at a very low spectral efficiency and ii) the proposed GFMA and FGMA can successfully guarantee high reliability and low latency at reasonably high spectral efficiency according to traffic class and QoS.

In addition, Fig. \ref{WMgain} shows the performance gain of employing the proposed waveform multiplexing in DL. Here, to clearly show the advantage of the proposed scheme, a single transmit antenna is instead assumed for each base station. The upper bound (assuming ideally controlled dynamic CP lengths, optimal OFDM parameters and ideal filter characteristics by Genie) on the spectral efficiency distribution of the proposed waveform multiplexing is shown and compared with the case of conventional LTE-based multiband OFDM. Here, the overall performance gain can be as high of 1.67 times, in which the two gains from selecting the ideal waveform on each subband according to OOBE characteristics and latency requirements and from selecting the optimal CP length and OFDM parameters are both meaningful in a realistic scenario. 

Although it might be too optimistic, a direct combination of the results in Figs. \ref{CDD} and \ref{WMgain} with those in Fig. \ref{GFMAPER} may anticipate that further gains in spectral efficiency with respect to those shown in Fig. \ref{GFMAPER} (up to 100\%) can be obtained by combining waveform multiplexing and CDD-SDMA with the proposed multiple access schemes.

\section{Conclusion}
In this paper, novel URLLC techniques were introduced for realizing Tactile Internet services in realistic environments. The traffic characteristics and required QoS of typical URLLC (or Tactile Internet) services in literature were summarized and classified from the perspective of designing the PHY and MAC layers of a cellular system. Investigations on typical traffic in typical use-cases justified the necessity of defining new user states and devising protocols for RRC connection according to latency requirements, multiplexing of multiple access schemes over radio resources to meet a variety of different traffic characteristics and QoS of URLLC services, and the development of latency-optimal radio resource management strategies to maximize the spectral efficiency while guaranteeing the latency and reliability requirements.

This paper proposed two additional user states aimed for low latency and devised the corresponding protocols and radio resource allocation strategies in detail. Further, a realistic map-based SLS approach was proposed based on a refined digital map construction, a realistic node distribution scenario, data collection via a ray-tracing tool, and the corresponding deterministic or stochastic 3D channel model.
Simulation results showed that the proposed schemes are promising for supporting URLLC services with high spectral efficiency while guaranteeing latency and reliability requirements.

To implement the proposed protocols and multiple access schemes in a spectrally efficient way, more PHY technologies on waveform multiplexing and synchronization strategy, channel codes for low processing delay and high reliability, and a novel DL/UL MU-MIMO concept combining interference alignment for a practical full-duplex cellular communication were further introduced, where each of them can provide significant performance improvement, even when incorporated with others, which encourages further efforts to substantiate the proposed work. 

\bibliographystyle{IEEEtran}
\bibliography{IEEEabrv,bibURLLC}

\begin{thebibliography}{100}
\providecommand{\url}[1]{#1}
\csname url@rmstyle\endcsname
\providecommand{\newblock}{\relax}
\providecommand{\bibinfo}[2]{#2}
\providecommand\BIBentrySTDinterwordspacing{\spaceskip=0pt\relax}
\providecommand\BIBentryALTinterwordstretchfactor{4}
\providecommand\BIBentryALTinterwordspacing{\spaceskip=\fontdimen2\font plus
\BIBentryALTinterwordstretchfactor\fontdimen3\font minus
  \fontdimen4\font\relax}
\providecommand\BIBforeignlanguage[2]{{%
\expandafter\ifx\csname l@#1\endcsname\relax
\typeout{** WARNING: IEEEtran.bst: No hyphenation pattern has been}%
\typeout{** loaded for the language `#1'. Using the pattern for}%
\typeout{** the default language instead.}%
\else
\language=\csname l@#1\endcsname
\fi
#2}}

\bibitem{IMT2020}
\BIBentryALTinterwordspacing
\emph{Framework and overall objectives of the future development of {IMT} for
  2020 and beyond}, Rec. M.2083, ITU-R, Sept. 2015. [Online]. Available:
  \url{https://www.itu.int/rec/R-REC-M.2083}
\BIBentrySTDinterwordspacing

\bibitem{ehealth}
\BIBentryALTinterwordspacing
\emph{{5G and e-health}}, White Paper, {5GPPP Association}, Oct. 2015.
  [Online]. Available:
  \url{https://5gppp.eu/wp-content/uploads/2016/02/5G-PPP-White-Paper-on-eHealth-Vertical-Sector.pdf}
\BIBentrySTDinterwordspacing

\bibitem{automation}
\BIBentryALTinterwordspacing
\emph{{5G automotive vision}}, White Paper, {5GPPP Association}, Oct. 2015.
  [Online]. Available:
  \url{https://5gppp.eu/wp-content/uploads/2014/02/5G-PPP-White-Paper-on-Automotive-Vertical-Sectors.pdf}
\BIBentrySTDinterwordspacing

\bibitem{industry}
\BIBentryALTinterwordspacing
\emph{{5G} and the Factories of the Future}, White Paper, {5GPPP Association},
  Oct. 2015. [Online]. Available:
  \url{https://5g-ppp.eu/wp-content/uploads/2014/02/5G-PPP-White-Paper-on-Factories-of-the-Future-Vertical-Sector.pdf}
\BIBentrySTDinterwordspacing

\bibitem{ind2}
M.~Luvisotto, Z.~Pang, and D.~Dzung, ``Ultra high performance wireless control
  for critical applications: Challenges and directions,'' \emph{{IEEE} Trans.
  Ind. Informat.}, vol.~13, no.~3, pp. 1448--1459, June 2017.

\bibitem{nokia}
\BIBentryALTinterwordspacing
\emph{{5G} for Mission Critical Communication}, White Paper, {Nokia}, 2016.
  [Online]. Available: \url{https://networks.nokia.com/innovation/5g}
\BIBentrySTDinterwordspacing

\bibitem{VKKV}
V.~Prasad, \emph{et~al.}, \emph{Tactile Usecase Summary}, VKKV-17, {IEEE}
  Standard P1918.1 Tactile Internet, Mar. 2017.

\bibitem{PHCR}
V.~Prasad, \emph{et~al.}, \emph{Consolidated Usecase for the {Tactile
  Internet}}, PHCR-17-4, {IEEE} Standard P1918.1 Tactile Internet, May 2017.

\bibitem{telesurgery1}
P.~Dario, B.~Hannaford, and A.~Menciassi, ``Smart surgical tools and augmenting
  devices,'' \emph{{IEEE} Trans. Robot. Autom.}, vol.~19, no.~5, pp. 782--792,
  Oct. 2003.

\bibitem{telesurgery2}
J.~Arata, \emph{et~al.}, ``Impact of network time-delay and force feedback on
  tele-surgery,'' \emph{International Journal of Computer Assisted Radiology
  and Surgery}, vol.~3, no.~3, pp. 371--378, Sept. 2008.

\bibitem{Fettweis}
G.~Fettweis and S.~Alamouti, ``{5G}: Personal mobile internet beyond what
  cellular did to telephony,'' \emph{{IEEE} Commun. Mag.}, vol.~52, no.~2, pp.
  140--145, Feb. 2014.

\bibitem{Steinbach}
E.~Steinbach and Q.~Liu, \emph{{Use case on teleoperation over the Tactile
  Internet}}, SL-16-3-r0, {IEEE} Standard P1918.1 Tactile Internet, July 2016.

\bibitem{3gpplow}
{3GPP}, ``Study on latency reduction techniques for {LTE} ({Release 14}),''
  3GPP TSG RAN, TR 36.881, May 2016.

\bibitem{Ericsson}
\emph{On {URLLC} design principles}, 3GPP TSG RAN WG1 Meeting \#86bis,
  R1-1609634, {Ericsson}, Oct. 2016.

\bibitem{IntelCorp}
\emph{Discussion on {URLLC} design aspects}, 3GPP TSG RAN WG1 Meeting \#86bis,
  R1-1610366, {Intel Corporation}, Oct. 2016.

\bibitem{Qualcomm}
\emph{{URLLC} numerology and frame structure design}, 3GPP TSG RAN WG1 Meeting
  \#86bis, R1-1610123, {Qualcomm}, Oct. 2016.

\bibitem{NRScenario}
3GPP, ``Study on scenarios and requirements for next generation access
  technologies ({Release 14}),'' 3GPP TSG RAN, TR 38.913, May 2017.

\bibitem{NRPHY}
3GPP, ``Study on new radio access technology; physical layer aspects ({Release
  14}),'' 3GPP TSG RAN, TR 36.802, Mar. 2017.

\bibitem{NRProtocol}
3GPP, ``Study on new radio access technology; radio interface protocol aspects
  ({Release 14}),'' 3GPP TSG RAN, TR 36.881, Mar. 2017.

\bibitem{DLmux1}
\emph{{UL URLLC} multiplexing considerations}, 3GPP TSG RAN WG1 Meeting \#87,
  R1-1611657, {Huawei} and {Hsillicon}, Nov. 2016.

\bibitem{DLmux2}
\emph{Scheduling and support for service multiplexing}, 3GPP TSG RAN WG1
  Meeting \#87, R1-1612316, {InterDigital Communications}, Nov. 2016.

\bibitem{DLmux3}
\emph{Multiplexing of {eMBB} and {URLLC} frame structures}, 3GPP TSG RAN WG1
  Meeting \#87, R1-1612316, {ZTE Corporation}, Nov. 2016.

\bibitem{gfma1}
\emph{{NR} Paging Design}, 3GPP TSG RAN WG1 Meeting \#87, R1-1611689, {LG
  Electronics}, Nov. 2016.

\bibitem{gfma2}
\emph{Grant-free to grant-based switching for {URLLC}}, 3GPP TSG RAN WG1
  Meeting \#87, R1-1611253, {Nokia and Alcatel-Lucent Shanghai Bell}, Nov.
  2016.

\bibitem{gfma3}
\emph{Cyclic delay-{Doppler} shifted {M}-Sequences for New Radio {PRACH}}, 3GPP
  TSG RAN WG1 Meeting \#87, R1-1611296, {Nokia and Alcatel-Lucent Shanghai
  Bell}, Nov. 2016.

\bibitem{LTEbook}
E.~Dahlman, S.~Parkvall, and J.~Skold, \emph{{4G}: {LTE/LTE-Advanced} for
  Mobile Broadband}, 1st~ed.\hskip 1em plus 0.5em minus 0.4em\relax Amsterdam:
  Academic Press, 2011.

\bibitem{diver}
\emph{Control channel design for {URLLC}}, 3GPP TSG RAN WG1 Meeting \#87,
  R1-1611221, {Huawei and Hsillicon}, Nov. 2016.

\bibitem{harq}
\emph{{HARQ} design for uplink grant-free transmission}, 3GPP TSG RAN WG1
  Meeting 90bis, R1-1717857, {Lenovo and Motorola Mobility}, Oct. 2017.

\bibitem{harq2}
J.~Yeo, \emph{et~al.}, ``Partial retransmission scheme for {HARQ} enhancement
  in {5G} wireless communications,'' in \emph{Proc. 4th Int. Workshop on URLLC
  in IEEE Globecom 2017}, Dec. 2017.

\bibitem{MarzettaNoncooperative}
T.~L. Marzetta, ``Noncooperative cellular wireless with unlimited numbers of
  base station antennas,'' \emph{{IEEE} Trans. Wireless Commun.}, vol.~9,
  no.~11, pp. 3590--3600, Nov. 2010.

\bibitem{Lim2015}
Y.-G. Lim, C.-B. Chae, and G.~Caire, ``Performance analysis of massive {MIMO}
  for cell-boundary users,'' \emph{{IEEE} Trans. Wireless Commun.}, vol.~14,
  no.~12, pp. 6827--6842, Dec. 2015.

\bibitem{GE}
G.~Hasslinger and O.~Hohlfeld, ``The {Gilbert-Elliott} model for packet loss in
  real time services on the {Internet},'' in \emph{14th GI/ITG Conference -
  Measurement, Modelling and Evalutation of Computer and Communication
  Systems}, Mar. 2008.

\bibitem{IntelIVR}
\BIBentryALTinterwordspacing
S.~Michalak, \emph{Guidelines for Immersive Virtual Reality Experiences}, White
  Paper, July 2017. [Online]. Available:
  \url{https://software.intel.com/en-us/articles/guidelines-for-immersive-virtual-reality-experiences}
\BIBentrySTDinterwordspacing

\bibitem{QualIVR}
\BIBentryALTinterwordspacing
\emph{Making Immersive Virtual Reality Possible in Mobile}, White Paper,
  Qualcomm, Apr. 2016. [Online]. Available:
  \url{https://www.qualcomm.com/documents/whitepaper-making-immersive-virtual-reality-possible-mobile}
\BIBentrySTDinterwordspacing

\bibitem{HuaweiIVR}
\BIBentryALTinterwordspacing
\emph{Virtual Reality/Augumented Reality White Paper}, {CAICT and Huawei},
  Sept. 2017. [Online]. Available:
  \url{http://www-file.huawei.com/-/media/CORPORATE/PDF/ilab/vr-ar-en.pdf}
\BIBentrySTDinterwordspacing

\bibitem{Tele.ProcIEEE}
S.~Hirche and M.~Buss, ``Human-oriented control for haptic teleoperation,''
  \emph{Proc. {IEEE}}, vol. 100, no.~3, pp. 623--647, Mar. 2012.

\bibitem{robocup}
M.~Asada, ``A report on {RoboCup} 2017 [competitions],'' \emph{{IEEE} Robot.
  Autom. Mag.}, vol.~24, no.~4, pp. 21--23, Dec. 2017.

\bibitem{MDohlerNews}
M.~Dohler, \emph{Will the {Tactile Internet} Globalize Your Skill Set?}, {IEEE}
  ComSoc Technology News, Jan. 2017.

\bibitem{UAV1}
D.~Lee, \emph{et~al.}, ``Semiautonomous haptic teleoperation control
  architecture of multiple unmanned aerial vehicles,'' \emph{{IEEE/ASME} Trans.
  Mechatronics}, vol.~18, no.~4, pp. 1334--1345, Aug. 2013.

\bibitem{IC1}
S.~A. Hossain, A.~S. M.~M. Rahman, and A.~E. Saddik, ``Measurements of
  multimodal approach to haptic interaction in second life interpersonal
  communication system,'' \emph{{IEEE} Trans. Instrum. Meas.}, vol.~60, no.~11,
  pp. 3547--3558, Nov. 2011.

\bibitem{IC2}
L.~philippe Morency, ``Modeling human communication dynamics [social
  sciences],'' \emph{{IEEE} Signal Process. Mag.}, vol.~27, no.~5, pp.
  112--116, Sept. 2010.

\bibitem{KimGFMA}
J.~Kim, \emph{et~al.}, ``Grant-free multiple access for ultra-reliable
  low-latency communications in a large-scale antenna system,'' in \emph{Proc.
  2016 Int. Conf. on Information and Communication Technology Convergence
  (ICTC)}, Oct. 2016, pp. 466--470.

\bibitem{RRCinactive_Nokia1}
S.~Hailu and M.~Saily, ``Hybrid paging and location tracking scheme for
  inactive {5G} {UEs},'' in \emph{2017 European Conference on Networks and
  Communications ({EuCNC})}.\hskip 1em plus 0.5em minus 0.4em\relax {IEEE},
  June 2017.

\bibitem{RRCinactive_Nokia2}
I.~L.~D. Silva, \emph{et~al.}, ``A novel state model for {5G} radio access
  networks,'' in \emph{2016 {IEEE} International Conference on Communications
  Workshops ({ICC})}.\hskip 1em plus 0.5em minus 0.4em\relax {IEEE}, May 2016.

\bibitem{RRCinactive_Nokia3}
\emph{Discussion of {RRC} States in {NR}}, 3GPP TSG RAN WG2 Meeting \#94,
  R2-163441, {Nokia and Alcatel-Lucent Shanghai Bell}, May 2016.

\bibitem{RRCinactive_5GPPP}
\BIBentryALTinterwordspacing
P.~Rugeland, \emph{et~al.}, \emph{Architectural enablers and concepts for
  mm-wave {RAN} integration}, White Paper, 5GPPP mmMAGIC, Mar. 2017. [Online].
  Available:
  \url{https://bscw.5g-mmmagic.eu/pub/bscw.cgi/d187833/mmMAGIC_Architectural_enablers_mmWave_integration.pdf}
\BIBentrySTDinterwordspacing

\bibitem{ChoiSPS}
K.~J. Choi and K.~S. Kim, ``Optimal semi-persistent uplink scheduling policy
  for large-scale antenna systems,'' \emph{{IEEE} Access}, vol.~5, pp.
  22\,902--22\,915, Oct. 2017.

\bibitem{Pop}
J.~J. Nielsen, R.~Liu, and P.~Popovski, ``Ultra-reliable low latency
  communication ({URLLC}) using interface diversity,'' \emph{arXiv preprint
  arXiv:1711.07771}, Nov. 2017.

\bibitem{HochwaldChannelHardening}
B.~M. Hochwald, T.~L. Marzetta, and V.~Tarokh, ``Multiple-antenna channel
  hargening and its implications for rate feedback and scheduling,''
  \emph{{IEEE} Trans. Inf. Theory}, vol.~50, no.~9, pp. 1893--1909, Sept. 2004.

\bibitem{GFMAACCESS}
J.~Kim and K.~S. Kim, ``Grant-free multiple access using {LSAS} for {URLLC}
  services,'' \emph{{IEEE} Access}, in preparation.

\bibitem{num}
\emph{Discussion on numerology multiplexing for supporting different service
  requirements}, 3GPP TSG RAN WG1 Meeting \#86bis, R1-1609145, {NEC}, Oct.
  2016.

\bibitem{IEEE_WCM_WM}
Y.-G. Lim, \emph{et~al.}, ``Waveform multiplexing for new radio: Numerology
  management and {3D} evaluation,'' \emph{{IEEE} Wireless Commun.}, to be
  published.

\bibitem{fulld}
M.~Chung, \emph{et~al.}, ``Prototyping real-time full duplex radios,''
  \emph{{IEEE} Commun. Mag.}, vol.~53, no.~9, pp. 56--63, Sept. 2015.

\bibitem{Chung2017}
M.~Chung, \emph{et~al.}, ``Compact full duplex {MIMO} radios in {D2D} underlaid
  cellular networks: From system design to prototype results,'' \emph{{IEEE}
  Access}, vol.~5, pp. 16\,601--16\,617, 2017.

\bibitem{fully}
R.~Li, \emph{et~al.}, ``Full-duplex cellular networks,'' \emph{{IEEE} Commun.
  Mag.}, vol.~55, no.~4, pp. 184--191, Apr. 2017.

\bibitem{IEEE_ComMag_WF}
A.~A. Zaidi, \emph{et~al.}, ``Waveform and numerology to support {5G} services
  and requirements,'' \emph{{IEEE} Commun. Mag.}, vol.~54, no.~11, pp. 90--98,
  Nov. 2016.

\bibitem{f-OFDMmux}
X.~Zhang, \emph{et~al.}, ``Filtered-{OFDM} - enabler for flexible waveform in
  the 5th generation cellular networks,'' in \emph{2015 {IEEE} Global
  Communications Conference ({GLOBECOM})}.\hskip 1em plus 0.5em minus
  0.4em\relax {IEEE}, Dec. 2015.

\bibitem{f-OFDM1}
J.~Abdoli, M.~Jia, and J.~Ma, ``Filtered {OFDM}: A new waveform for future
  wireless systems,'' in \emph{Proc. IEEE Int. Workshop on SPAWC}.\hskip 1em
  plus 0.5em minus 0.4em\relax {IEEE}, June 2015, pp. 66--70.

\bibitem{f-OFDM2}
\emph{f-{OFDM} scheme and filter design}, 3GPP TSG RAN WG1 Meeting \#85,
  R1-165425, Huawei and Hsillicon, May 2016.

\bibitem{GFDM1}
G.~Fettweis, M.~Krondorf, and S.~Bittner, ``{GFDM} - generalized frequency
  division multiplexing,'' in \emph{{VTC} Spring 2009 - {IEEE} 69th Vehicular
  Technology Conference}.\hskip 1em plus 0.5em minus 0.4em\relax {IEEE}, Apr.
  2009.

\bibitem{GFDM2}
N.~Michailow, \emph{et~al.}, ``Generalized frequency division multiplexing for
  5th generation cellular networks,'' \emph{{IEEE} Trans. Commun.}, vol.~62,
  no.~9, pp. 3045--3061, Sept. 2014.

\bibitem{FC-OFDM}
H.~Lin, ``Flexible configured {OFDM} for {5G} air interface,'' \emph{{IEEE}
  Access}, vol.~3, pp. 1861--1870, 2015.

\bibitem{sync_twc}
B.~Lim and Y.-C. Ko, ``{SIR} analysis of {OFDM} and {GFDM} waveforms with
  timing offset, {CFO}, and phase noise,'' \emph{{IEEE} Trans. Wireless
  Commun.}, vol.~16, no.~10, pp. 6979--6990, Oct. 2017.

\bibitem{sync_wcnc}
B.~Lim and Y.-C. Ko, ``Optimal receiver filter for {GFDM} with timing and
  frequency offsets in uplink multiuser systems,'' in \emph{2018 {IEEE}
  Wireless Communications and Networking Conference ({WCNC})}.\hskip 1em plus
  0.5em minus 0.4em\relax {IEEE}, Apr. 2018.

\bibitem{ARACA}
K.~J. Jeon and K.~S. Kim, ``Accumulate repeat accumulate check accumulate
  codes,'' \emph{{IEEE} Trans. Commun.}, vol.~65, no.~11, pp. 4585--4599, Nov.
  2017.

\bibitem{RCARACA}
K.~J. Jeon and K.~S. Kim, ``Rate-compatible {ARACA} codes,'' \emph{Electron.
  Lett.}, vol.~54, no.~6, pp. 398--400, Mar. 2018.

\bibitem{Morelli}
M.~Morelli, C.-C.~J. Kuo, and M.-O. Pun, ``Synchronization techniques for
  orthogonal frequency division multiple access ({OFDMA}): A tutorial review,''
  \emph{Proc. {IEEE}}, vol.~95, no.~7, pp. 1394--1427, July 2007.

\bibitem{Gaspar}
I.~S. Gaspar, \emph{et~al.}, ``A synchronization technique for generalized
  frequency division multiplexing,'' \emph{EURASIP Journal on Advances in
  Signal Processing}, vol. 2014, no.~1, p.~67, May 2014.

\bibitem{Morelliuplink}
M.~Morelli, ``Timing and frequency synchronization for the uplink of an {OFDMA}
  system,'' \emph{{IEEE} Trans. Commun.}, vol.~52, no.~2, pp. 296--306, Feb.
  2004.

\bibitem{Beek}
J.-J. van~de Beek, \emph{et~al.}, ``A time and frequency synchronization scheme
  for multiuser {OFDM},'' \emph{{IEEE} J. Sel. Areas Commun.}, vol.~17, no.~11,
  pp. 1900--1914, 1999.

\bibitem{Manohar}
S.~Manohar, \emph{et~al.}, ``Cancellation of multiuser interference due to
  carrier frequency offsets in uplink {OFDMA},'' \emph{{IEEE} Trans. Wireless
  Commun.}, vol.~6, no.~7, pp. 2560--2571, July 2007.

\bibitem{Huang}
D.~Huang and K.~Letaief, ``An interference-cancellation scheme for carrier
  frequency offsets correction in {OFDMA} systems,'' \emph{{IEEE} Trans.
  Commun.}, vol.~53, no.~7, pp. 1155--1165, July 2005.

\bibitem{5Gcc}
\emph{Evolved Universal Terrestrial Radio Access ({E-UTRA}); Multiplexing and
  channel coding}, 3GPP TS 36.212, Rev. 14.4.0, Sept. 2017.

\bibitem{Error}
T.~Richardson, ``{Error-floors of LDPC codes},'' in \emph{Proc. of the 41th
  Annual Allerton Conf. on Commun., Control and Computing}, Sept. 2003.

\bibitem{ARA}
D.~Divsalar, S.~Dolinar, and J.~Thorpe,
  ``Accumulate-repeat-accumulate-accumulate-codes,'' in \emph{{IEEE} 60th
  Vehicular Technology Conference ({VTC}2004-Fall)}, Sept. 2004, pp.
  1271--1275.

\bibitem{ARJA}
D.~Divsalar, \emph{et~al.}, ``Protograph based {LDPC} codes with minimum
  distance linearly growing with block size,'' in \emph{2005 {IEEE} Global
  Communications Conference ({GLOBECOM})}, Nov. 2005.

\bibitem{SSS}
\emph{Preliminary evaluation results on new channel coding scheme for {NR}},
  3GPP TSG RAN WG1 Meeting \#85, R1-164812, {Samsung}, May 2016.

\bibitem{RCSL}
T.~V. Nguyen and A.~Nosratinia, ``Rate-compatible short-length protograph
  {LDPC} codes,'' \emph{{IEEE} Commun. Lett.}, vol.~17, no.~5, pp. 948--951,
  May 2013.

\bibitem{PBRL}
T.-Y. Chen, \emph{et~al.}, ``Protograph-based raptor-like {LDPC} codes,''
  \emph{{IEEE} Trans. Commun.}, vol.~63, no.~5, pp. 1522--1532, May 2015.

\bibitem{ARACA2}
K.~J. Jeon and K.~S. Kim, ``High-speed encoder/decoder structure for {ARACA}
  codes,'' \emph{{IEEE} Trans. Commun.}, in preparation.

\bibitem{Cadambe:08}
V.~Cadambe and S.~Jafar, ``Interference alignment and degrees of freedom of the
  {$K$}-user interference channel,'' \emph{{IEEE} Trans. Inf. Theory}, vol.~54,
  no.~8, pp. 3425--3441, Aug. 2008.

\bibitem{Yang:17}
M.~Yang, S.-W. Jeon, and D.~K. Kim, ``Degrees of freedom of full-duplex
  cellular networks with reconfigurable antennas at base station,''
  \emph{{IEEE} Trans. Wireless Commun.}, vol.~16, no.~4, pp. 2314--2326, Apr.
  2017.

\bibitem{3DSCM}
3GPP, ``Study on {3D} channel model for {LTE},'' 3GPP TSG RAN, TR 36.873, June
  2015.

\bibitem{MapCh_Rapa}
S.~Seidel and T.~Rappaport, ``Site-specific propagation prediction for wireless
  in-building personal communication system design,'' \emph{{IEEE} Trans. Veh.
  Technol.}, vol.~43, no.~4, pp. 879--891, 1994.

\bibitem{Wise_Mag}
V.~Erceg, \emph{et~al.}, ``Comparisons of a computer-based propagation
  prediction tool with experimental data collected in urban microcellular
  environments,'' \emph{{IEEE} J. Sel. Areas Commun.}, vol.~15, no.~4, pp.
  677--684, May 1997.

\bibitem{MapCh_Heath}
R.~Bhagavatula, \emph{et~al.}, ``Sizing up {MIMO} arrays,'' \emph{{IEEE} Veh.
  Technol. Mag.}, vol.~3, no.~4, pp. 31--38, Dec. 2008.

\bibitem{MapCh_3GPP}
{3GPP}, ``Study on channel model for frequency spectrum above 6 {GHz},'' 3GPP
  TSG RAN, TR 38.900, June 2018.

\bibitem{Map_METIS}
\emph{{METIS} Channel Models}, ICT-317669 METIS Project deliverable D1.4 v.3,
  METIS, June 2015.

\bibitem{Map_METIS_Mag}
J.~Medbo, \emph{et~al.}, ``Radio propagation modeling for {5G} mobile and
  wireless communications,'' \emph{{IEEE} Commun. Mag.}, vol.~54, no.~6, pp.
  144--151, June 2016.

\bibitem{Map_Lim}
Y.-G. Lim, \emph{et~al.}, ``Map-based millimeter-wave channel models: An
  overview, guidelines, and data,'' \emph{arXiv preprint arXiv: 1711.09052},
  Nov. 2017.

\bibitem{Jang_Smallcell}
J.~Jang, \emph{et~al.}, ``Smart small cell with hybrid beamforming for {5G}:
  Theoretical feasibility and prototype results,'' \emph{{IEEE} Wireless
  Commun.}, vol.~23, no.~6, pp. 124--131, Dec. 2016.

\bibitem{Oh_DP}
T.~Oh, \emph{et~al.}, ``Dual-polarization slot antenna with high
  cross-polarization discrimination for indoor small-cell {MIMO} systems,''
  \emph{{IEEE} Antennas Wireless Propag. Lett.}, vol.~14, pp. 374--377, Feb.
  2015.

\bibitem{Kwon_TMTT_Lens}
T.~Kwon, \emph{et~al.}, ``{RF} lens-embedded massive {MIMO} systems:
  Fabrication issues and codebook design,'' \emph{{IEEE} Trans. Microw. Theory
  Techn.}, vol.~64, no.~7, pp. 2256--2271, July 2016.

\bibitem{Sim_FD}
M.~S. Sim, \emph{et~al.}, ``Nonlinear self-interference cancellation for
  full-duplex radios: From link-level and system-level performance
  perspectives,'' \emph{{IEEE} Commun. Mag.}, vol.~55, no.~9, pp. 158--167,
  Sept. 2017.

\bibitem{JSAC_mmWaveRT}
B.~Ai, \emph{et~al.}, ``On indoor millimeter wave massive {MIMO} channels:
  Measurement and simulation,'' \emph{{IEEE} J. Sel. Areas Commun.}, vol.~35,
  no.~7, pp. 1678--1690, July 2017.

\bibitem{Access_mmWaveRT}
V.~Degli-Esposti, \emph{et~al.}, ``Ray-tracing-based mm-wave beamforming
  assessment,'' \emph{{IEEE} Access}, vol.~2, pp. 1314--1325, 2014.

\bibitem{Kim_NOMA}
H.~Kim, \emph{et~al.}, ``Multiple access for {5G} new radio: Categorization,
  evaluation, and challenges,'' \emph{arXiv preprint arXiv: 1703.09042}, Mar.
  2017.

\bibitem{Lim_DP}
Y.-G. Lim, \emph{et~al.}, ``Relationship between cross-polarization
  discrimination ({XPD}) and spatial correlation in indoor small-cell {MIMO}
  systems,'' \emph{{IEEE} Wireless Commun. Lett.}, vol.~7, no.~4, pp. 654--657,
  Aug. 2018.

\bibitem{Cho2018}
Y.~J. Cho, \emph{et~al.}, ``{RF} lens-embedded antenna array for {mmWave}
  {MIMO}: Design and performance,'' \emph{{IEEE} Commun. Mag.}, vol.~56, no.~7,
  pp. 42--48, July 2018.

\bibitem{KICS_RYU}
K.~L. Ryu, \emph{et~al.}, ``{MIESM} based system-level simulator for {3D}
  realistic {MIMO} channel,'' in \emph{2018 KICS Winter Conference}.\hskip 1em
  plus 0.5em minus 0.4em\relax {KICS}, Jan. 2018.

\end{thebibliography}

\begin{IEEEbiography}[{\includegraphics[width=1in,height=1.25in,clip,keepaspectratio]{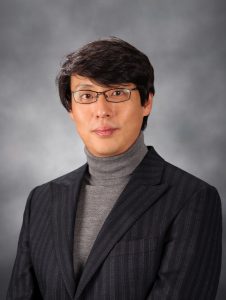}}]{Kwang~Soon~Kim}
(S'95--M'99--SM'04) received the B.S. (summa cum laude), M.S.E., and Ph.D. degrees in Electrical Engineering from Korea Advanced Institute of Science and Technology (KAIST), Daejeon, Korea, in February 1994, February 1996, and February 1999, respectively.
From March 1999 to March 2000, he was with the Department of Electrical and Computer Engineering, University of California at San Diego, La Jolla, CA, U.S.A., as a Postdoctoral Researcher. From April 2000 to February 2004, he was with the Mobile Telecommunication Research Laboratory, Electronics and Telecommunication Research Institute, Daejeon, Korea as a Senior Member of Research Staff.
Since March 2004, he has been with the Department of Electrical and Electronic Engineering, Yonsei University, Seoul, Korea, now is a Professor. Prof. Kim is a Senior Member of the IEEE, served as an Editor of the Journal of the Korean Institute of Communications and Information Sciences (KICS) from 2006 to 2012, as the Editor-in-Chief of the journal of KICS from 2013 to 2016, as an Editor of the Journal of Communications and Networks (JCN) since 2008, as an Editor of the IEEE Transactions on Wireless Communications from 2009 to 2014.
He was a recipient of the Postdoctoral Fellowship from Korea Science and Engineering Foundation (KOSEF) in 1999. He received the Outstanding Researcher Award from Electronics and Telecommunication Research Institute (ETRI) in 2002, the Jack Neubauer Memorial Award (Best system paper award, IEEE Transactions on Vehicular Technology) from IEEE Vehicular Technology Society in 2008, and LG R\&D Award: Industry-Academic Cooperation Prize, LG Electronics, 2013. His research interests are generally in signal processing, communication theory, information theory, and stochastic geometry applied to wireless heterogeneous cellular networks, wireless local area networks, wireless D2D networks and wireless ad doc networks, and are focused on the new radio access technologies for 5G, recently.
\end{IEEEbiography}

\begin{IEEEbiography}[{\includegraphics[width=1in,height=1.25in,clip,keepaspectratio]{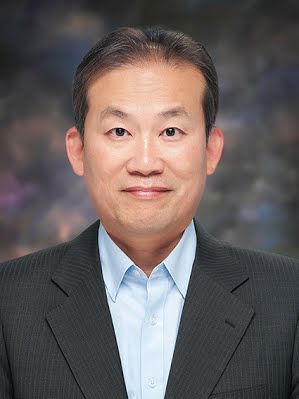}}]{Dong~Ku~Kim}
(SM'15)  received his B.S. from Korea Aerospace University in 1983, and his M.S. and Ph.D. from the University of Southern California, Los Angeles, in 1985 and 1992, respectively.
He worked on CDMA systems in the cellular infrastructure group of Motorola at Fort Worth, Texas, in 1992.
He has been a professor in the School of Electrical and Electronic Engineering, Yonsei University since 1994 and a Vice Chair of the Executive Committee of the 5G Forum since 2013.
Currently, he is a Vice President for Academic Research Affairs of the KICS.
He received the Minister Award for Distinguished Service for ICT R\&D from the MSIP in 2013, the Award of Excellence in leadership of 100 Leading Core Technologies for Korea 2020 from the NAEK in 2013, and the Dr. Irwin Jacobs Academic Achievement Award 2016 from Qualcomm and KICS.
\end{IEEEbiography}

\begin{IEEEbiography}[{\includegraphics[width=1in,height=1.25in,clip,keepaspectratio]{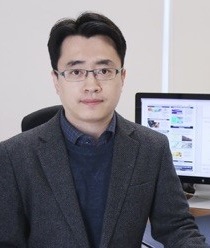}}]{Chan-Byoung~Chae}
(SM'12) is the Underwood Distinguished Professor at Yonsei University. Before joining Yonsei, he was with Bell Laboratories and Harvard University. He received his Ph.D. from the University of Texas at Austin (2008). He was the recipient of the IEEE INFOCOM Best Demo Award (2015), the IEEE SPMag Best Paper Award (2013), the IEEE ComSoc Outstanding Young Researcher Award (2012), and the IEEE Daniel Noble Fellowship Award (2008). He serves/has served as an Editor for IEEE TWC, IEEE Communications Magazine, IEEE WCL, IEEE JSAC, and IEEE T-MBMC.
\end{IEEEbiography}

\begin{IEEEbiography}[{\includegraphics[width=1in,height=1.25in,clip,keepaspectratio]{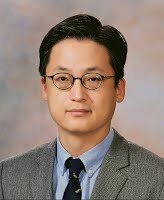}}]{Sunghyun~Choi}
(S'96--M'00--SM'05--F'14) is a professor at the Department of Electrical and Computer Engineering, Seoul National University (SNU), Korea. He received his B.S. (summa cum laude) and M.S. degrees from Korea Advanced Institute of Science and Technology in 1992 and 1994, respectively, and received Ph.D. from The University of Michigan, Ann Arbor in 1999. He co-authored over 230 technical papers and holds over 160 patents in the areas of wireless/mobile networks and communications.  He served on the editorial boards of IEEE Transactions on Mobile Computing, IEEE Transactions on Wireless Communications, and IEEE Wireless Communications Magazine. 
\end{IEEEbiography}

\begin{IEEEbiography}[{\includegraphics[width=1in,height=1.25in,clip,keepaspectratio]{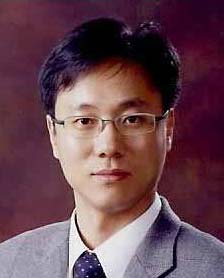}}]{Young-Chai~Ko}
(S'97-–M'01-–SM'06) received the B.Sc. degree in electrical and telecommunication engineering from Hanyang University, Seoul, South Korea, and the M.S.E.E. and Ph.D. degrees in electrical engineering from the University of Minnesota, Minneapolis, MN, in 1999 and 2001, respectively. He was with Novatel Wireless as a Research Scientist in 2001. In 2001, he joined the Wireless Center, Texas Instruments, Inc., San Diego, CA, as a Senior Engineer. He is currently a Professor with the School of Electrical Engineering, Korea University. His current research interests are the performance analysis and the design of wireless communication systems.
\end{IEEEbiography}

\begin{IEEEbiography}[{\includegraphics[width=1in,height=1.25in,clip,keepaspectratio]{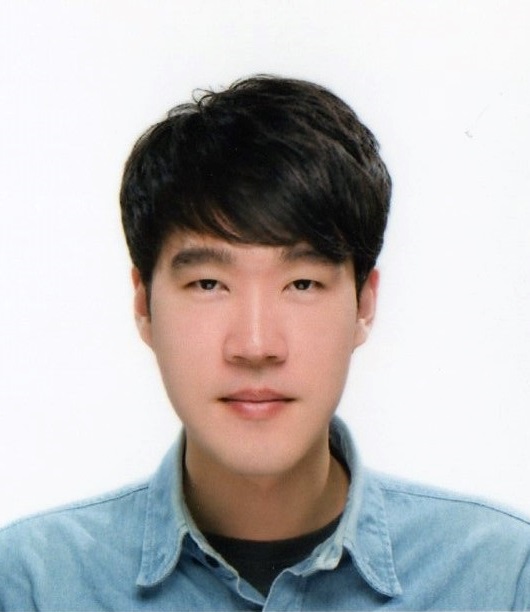}}]{Jonghyun~Kim}
(S'15) received the B.S. degree in electrical and electronic engineering from Yonsei University, Seoul, Korea, in 2016, where he is currently pursuing the combined M.S. and Ph.D. degree with the School of Electrical and Electronic Engineering.
He has been involved in several national R\&D projects in Korea supported by ADD, NRF, and IITP since 2015.
From 2013 to 2014, he was with Qualcomm CDMA Technologies Korea.
His current research interests include MIMO waveform transceiver design and radio access technology for 5G and B5G communications.
\end{IEEEbiography}

\begin{IEEEbiography}[{\includegraphics[width=1in,height=1.25in,clip,keepaspectratio]{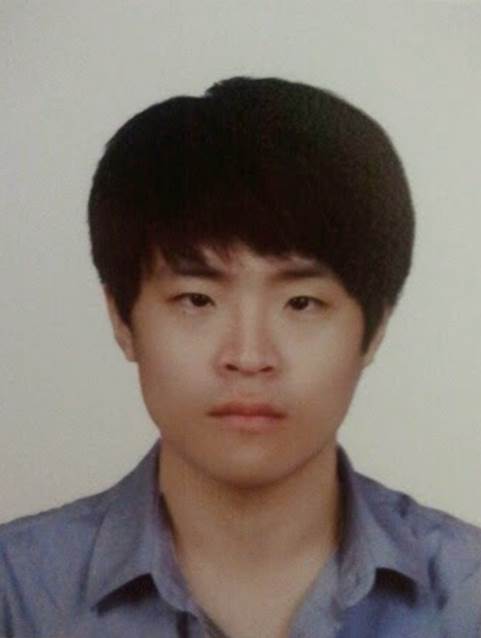}}]{{Yeon-Geun}~Lim}
(S'12) received his B.S. degree in Information and Communications Engineering from Sungkyunkwan University, Korea in 2012.
He is now with the School of Integrated Technology, Yonsei University and is working toward the Ph.D. degree.
He was the recipient of Samsung Humantech Paper Award (2018).
He has been involved in several industrial and national projects sponsored by Samsung Electronics, IITP, and KCA. 
His research interest includes massive MIMO, next-generation waveforms, full-duplex, mm-Wave technologies, and system level simulation for 5G networks.
\end{IEEEbiography}

\begin{IEEEbiography}[{\includegraphics[width=1in,height=1.25in,clip,keepaspectratio]{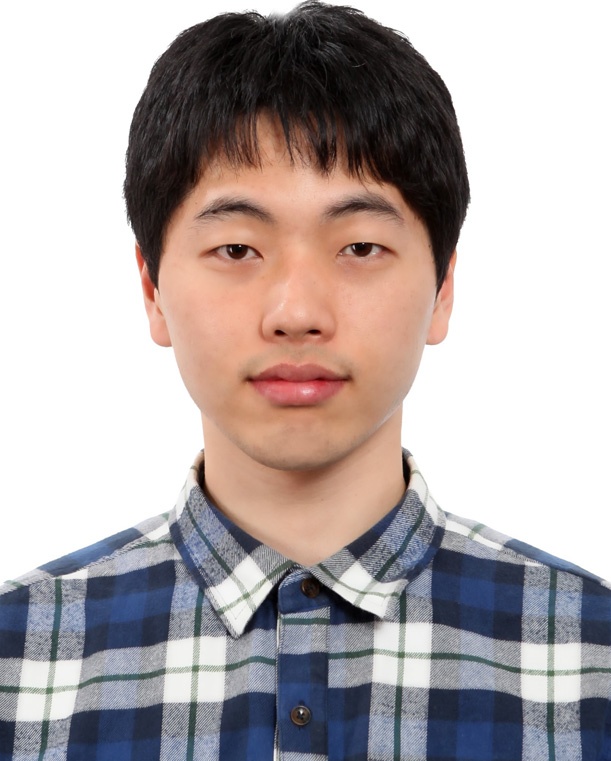}}]{Minho~Yang}
(S'13) received  his  B.S.  degree from the  School  of  Electrical  and  Electronic  Engineering at  Yonsei  University,  Seoul,  Korea,  in  2012.  
He is now working toward his Ph.D. degree at Yonsei University, Seoul, Korea.  
His research interests include network information theory and wireless communication systems.
\end{IEEEbiography}

\begin{IEEEbiography}[{\includegraphics[width=1in,height=1.25in,clip,keepaspectratio]{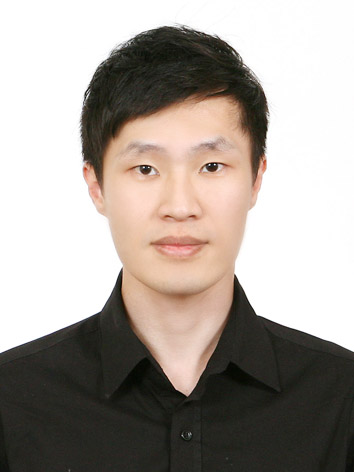}}]{Sundo~Kim}
(S'13) received the B.S. degree from Korea Advanced Institute of Science and Technology, Daejeon, Korea. He is currently working toward the Ph.D. degree with the Department of Electrical and Computer Engineering, Seoul National University. His research interests include latency reduction algorithm and low latency protocol development for fifth-generation networks.
\end{IEEEbiography}

\begin{IEEEbiography}[{\includegraphics[width=1in,height=1.25in,clip,keepaspectratio]{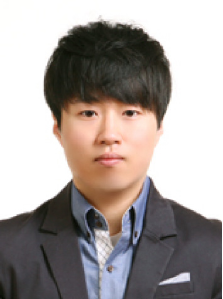}}]{Byungju~Lim}
received the B.S. and M.S. degrees in electrical engineering from Korea University, Seoul, Korea in 2015 and 2017, respectively.
He is currently working toward the Ph.D. degree in the School of Electrical Engineering at Korea University.
His current research interests include synchronization, multi-carrier systems, and signal processing.
\end{IEEEbiography}

\begin{IEEEbiography}[{\includegraphics[width=1in,height=1.25in,clip,keepaspectratio]{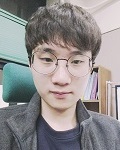}}]{Kwanghoon~Lee}
received the B.S. degree in electrical and electronic engineering from Yonsei University, Seoul, South Korea, in 2016, where he is currently pursuing the combined M.S. and Ph.D. degree with the School of Electrical and Electronic Engineering.
He has been involved in several national R\&D projects in Korea supported by ADD, NRF, and IITP since 2016.
His current research interests include radio resource management, mm-Wave beamforming, and realistic system-level simulation for V2X networks.
\end{IEEEbiography}

\begin{IEEEbiography}[{\includegraphics[width=1in,height=1.25in,clip,keepaspectratio]{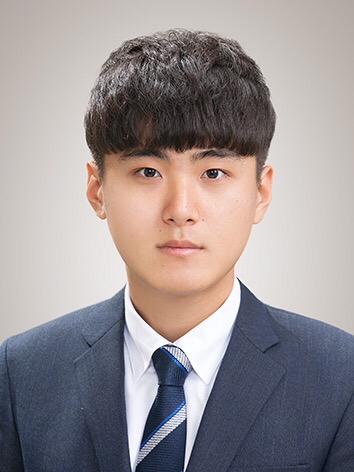}}]{Kyung~Lin~Ryu}
was born in Busan, South Korea, in 1995. He received the B.S. degrees in Electronic and Radio Engineering in 2017 from KyungHee University, Yongin, South Korea, and he is currently working toward the M.S. degree with the School of Electrical and Electronic Engineering from Yonsei University, Seoul.
He has been involved in several national R\&D projects in Korea supported by NRF and IITP since 2017.
His research interests include massive MIMO and grant-free multiple access.
\end{IEEEbiography}

\end{document}